\documentclass[12pt, a4paper]{article}

\usepackage[total={39pc, 55.5pc}, top=30mm, centering]{geometry}

\usepackage{setspace}
\usepackage{amsmath, amsthm, amssymb, amsfonts, mathrsfs, amsbsy, bm}

\usepackage{graphicx}
\usepackage{xcolor} 

\usepackage{algorithm}
\usepackage{algpseudocode}

\usepackage[nosort]{cite}

\usepackage[hidelinks]{hyperref}

\newcommand{\bx}{\mathbf{x}}
\newcommand{\by}{\mathbf{y}}

\newcommand{\bbC}{\mathbb{C}}
\newcommand{\bbR}{\mathbb{R}}

\newcommand{\calF}{\mathcal{F}}
\newcommand{\calJ}{\mathcal{J}}
\newcommand{\calK}{\mathcal{K}}

\newcommand{\calO}{\mathcal{O}}

\newcommand{\calR}{\mathcal{R}}
\newcommand{\calS}{\mathcal{S}}

\newcommand{\ReS}{\mathrm{Re}\,S}
\newcommand{\ImS}{\mathrm{Im}\,S}

\newcommand{\bra}[1]{\langle #1 |}
\newcommand{\ket}[1]{| #1 \rangle}
\newcommand{\vev}[1]{\langle #1 \rangle}

\newcommand{\braket}[2]{\langle #1 | #2 \rangle}

\newcommand{\tr}{\mathrm{tr}\,}
\newcommand{\re}{\mathrm{Re}\,}

\newcommand{\with}{{~~\mbox{with}~~}}

\makeatletter
\@addtoreset{equation}{section}

\def\@seccntformat#1{\csname the#1\endcsname.~~}
\makeatother

\begin{document}

\begin{titlepage} 

\renewcommand{\thefootnote}{\fnsymbol{footnote}}
\begin{flushright}
  KUNS-3065
\end{flushright}
\vspace*{1.0cm}

\begin{center}
{\Large \bf
Applying the Worldvolume Hybrid Monte Carlo method to the Hubbard model away from half filling
}

\vspace{1.0cm}

\centerline{
  {Masafumi Fukuma${}^1$}%
  \footnote{
    E-mail address: fukuma@gauge.scphys.kyoto-u.ac.jp
  }
  and 
  {Yusuke Namekawa${}^2$}%
  \footnote{
    E-mail address: namekawa@fukuyama-u.ac.jp
  }
}

\vskip 0.8cm
${}^1${\it Department of Physics, Kyoto University,
  Kyoto 606-8502, Japan}
\vskip 0.1cm
${}^2${\it Department of Computer Science, Fukuyama University,
  Hiroshima 729-0292, Japan}
\vskip 1.2cm

\end{center}

\begin{abstract}

The Worldvolume Hybrid Monte Carlo (WV-HMC) method [arXiv:2012.08468] is an efficient algorithm for addressing the numerical sign problem at moderate computational cost. It mitigates the sign problem while avoiding the ergodicity issues inherent in approaches based on Lefschetz thimbles. In this study, we apply WV-HMC to the two-dimensional Hubbard model doped away from half filling, which is known to suffer from a severe sign problem. We compute the number density and the energy density on lattices of size $6 \times 6$ and $8 \times 8$ at temperature $T/t = 1/6.4 \simeq 0.156$ and interaction strength $U/t = 8.0$, using Trotter number $N_t = 20$ (Trotter step $\epsilon = 0.32$). Our results demonstrate that WV-HMC remains effective even in parameter regimes where standard (non-thimble) determinant quantum Monte Carlo methods fail. In this work, fermion matrix inversions are performed using direct solvers, leading to a computational cost of $O(N^3)$, where $N$ denotes the number of degrees of freedom and is proportional to the spacetime lattice volume. An alternative algorithm employing pseudofermions and iterative solvers, which reduces the cost to $O(N^2)$ at the expense of careful parameter tuning, will be discussed in a separate publication.

\end{abstract}
\end{titlepage}

\pagestyle{empty}
\pagestyle{plain}

\tableofcontents
\setcounter{footnote}{0}

\section{Introduction}
\label{sec:intro}

The numerical sign problem arises in a variety of physically important systems, 
including quantum chromodynamics (QCD) at finite density, 
strongly correlated electron systems, 
and real-time dynamics of quantum many-body systems. 
Among recent efforts to develop versatile algorithms to address this problem, 
the Lefschetz thimble method 
\cite{Witten:2010cx,Cristoforetti:2012su,Cristoforetti:2013wha,
Fujii:2013sra,Fujii:2015bua,Fujii:2015vha,
Alexandru:2015xva,Alexandru:2015sua,Alexandru:2017lqr} 
has emerged as a promising approach. 
This method continuously deforms the original integration surface 
$\Sigma_0 = \bbR^N$ 
(with $N$ denoting the number of degrees of freedom) 
into a submanifold $\Sigma$ within the complexified space $\bbC^N$. 
The deformed surface asymptotically approaches a union of Lefschetz thimbles, 
on each of which the imaginary part of the action is constant, 
thereby suppressing phase fluctuations of the integrand 
and alleviating the sign problem.
However, the presence of infinitely high potential barriers 
between adjacent thimbles leads to an ergodicity problem.
This issue was resolved by performing (parallel) tempering 
with respect to the deformation parameter, 
as implemented in the tempered Lefschetz thimble (TLT) method 
\cite{Fukuma:2017fjq,Alexandru:2017oyw} 
(see also Ref.~\cite{Fukuma:2019wbv}). 
A principal drawback of this approach is its high computational cost, 
primarily due to the need to evaluate the Jacobian of the deformation 
at every exchange of configurations, 
in order to take into account 
the difference between the volume elements of the replicas. 

The \emph{Worldvolume Hybrid Monte Carlo} (WV-HMC) method 
\cite{Fukuma:2020fez} 
(see also Refs.~\cite{Fukuma:2021aoo,Fukuma:2023eru,
Fukuma:2025gya,Namekawa:2024ert,Fukuma:2025esu}) 
was then introduced to address this issue. 
In this algorithm, Hybrid Monte Carlo (HMC) updates are performed 
over a continuous union of deformed integration surfaces.
This region is referred to as the \emph{worldvolume} 
because it can be viewed as the orbit of the integration surface 
in the target space $\bbC^N$ 
(or as the orbit in $G^\bbC$ 
when the original integration space $\Sigma_0$ 
is a compact Lie group $G$ \cite{Fukuma:2025gya}). 
The sampling is performed over the tangent bundle of the worldvolume, 
which has a natural symplectic structure. 
We no longer need to compute the Jacobian in configuration generation 
because the phase-space volume element does not change 
provided that we employ a symplectic integrator 
(such as RATTLE \cite{Andersen:1983,Leimkuhler:1994}) 
in molecular dynamics.

The Hubbard model has long been regarded as one of the most important models 
in condensed matter physics, 
as it provides a minimal description of the essential competition 
between electron itinerancy and local Coulomb repulsion in solids. 
The model is also highly relevant to particle physics, 
because its bosonized form 
(obtained through the Hubbard--Stratonovich transformation) 
shares structural similarities with that in finite-density QCD.
However, the model suffers from a severe sign problem 
when it is doped away from half filling. 
A variety of numerical methods have been developed 
to address this issue, 
including variational Monte Carlo 
\cite{Yokoyama:1987,Sorella:2005,Yamaji:1998,Tahara:2008}, 
constrained-path auxiliary-field quantum Monte Carlo 
\cite{Zhang:1995zz,Zhang:1997}, 
as well as more recent approaches  
such as the Lefschetz thimble method 
\cite{Mukherjee:2014hsa,Ulybyshev:2017hbs,Ulybyshev:2019hfm,
Ulybyshev:2019fte,Ulybyshev:2022kxq,Ulybyshev:2024kdr}, 
the TLT method \cite{Fukuma:2019wbv}, 
tensor renormalization group techniques 
\cite{Akiyama:2021xxr,Akiyama:2021glo}, 
complex-valued neural networks \cite{Rodekamp:2022xpf}, 
constant path-integral contour shifts \cite{Gantgen:2023byf}, 
and equivariant normalizing flows \cite{Schuh:2025fky}. 

In this paper, 
we apply the WV-HMC method to the doped Hubbard model.
We compute the number and energy densities 
on $6 \times 6$ and $8 \times 8$ lattices 
at temperature $T/t = 1/6.4 \simeq 0.156$ 
and interaction strength $U/t = 8.0$ 
for various values of the chemical potential $\mu$, 
where $t$ denotes the hopping amplitude. 
The temporal direction is discretized using Trotter number $N_t = 20$ 
(corresponding to Trotter step $\epsilon = 0.32$).%
\footnote{ 
  The continuum limit $\epsilon\to 0$ is discussed 
  in a subsequent paper \cite{Fukuma:2025cxg}, 
  where a machinery is developed 
  to enable simulations on larger spacetime lattices.
} 
We compare our results with those obtained using ALF 
(Algorithms for Lattice Fermions) \cite{Bercx:2017pit,ALF:2020tyi}, 
a highly efficient, state-of-the-art, non-thimble 
determinant quantum Monte Carlo (DQMC) framework 
widely used in the condensed-matter community. 
In our implementation, 
we employ direct solvers for inverting fermion matrices, 
which leads to a computational cost scaling as $O(N^3)$, 
where $N$ is the number of degrees of freedom 
and is proportional to the spacetime lattice volume. 
An alternative formulation using pseudofermions and iterative solvers,  
which reduces the cost to $O(N^2)$  
at the expense of careful parameter tuning, 
will be presented in a separate publication. 

This paper is organized as follows.
Section~\ref{sec:wv-hmc} reviews the WV-HMC algorithm, 
in which the computation of observables are reduced 
to phase-space integrals over the tangent bundle of the worldvolume, 
which carries a natural symplectic structure.  
Section~\ref{sec:path_integral} 
presents a path-integral formulation of the Hubbard model 
that is well suited for WV-HMC simulations. 
Section~\ref{sec:app} provides the explicit formulas 
used in applying WV-HMC to the Hubbard model. 
Section~\ref{sec:results1} reports numerical results 
for the one-dimensional doped Hubbard model at high temperature, 
confirming that WV-HMC reproduces the exact values with high accuracy. 
Section~\ref{sec:results2} presents simulations 
of the two-dimensional doped Hubbard model at low temperature, 
where the computational cost is shown to scale as $O(N^3)$. 
The number and energy densities are estimated 
on $6 \times 6$ and $8 \times 8$ spatial lattices 
using the physical parameters specified above, 
and are shown to yield statistically controlled results 
even in parameter regions 
where conventional DQMC fails due to severe sign problems. 
Section~\ref{sec:conclusions} is devoted to conclusions and outlook, 
with particular emphasis 
on extending the simulations to larger spacetime volumes. 
Preliminary results of this study were reported 
at Lattice conferences~\cite{Namekawa:2024ert,Fukuma:2025esu}.

\section{Worldvolume Hybrid Monte Carlo (WV-HMC)}
\label{sec:wv-hmc}

In this section, 
we give a brief introduction to the WV-HMC method. 
See Refs.~\cite{Fukuma:2020fez,Fukuma:2021aoo,Fukuma:2023eru,Fukuma:2025gya} 
for details.

Our aim is to numerically estimate 
the expectation value of an observable $\calO(x)$ 
defined by a path integral over the configuration space $\mathbb{R}^N = \{x\}$,
\begin{align}
  \vev{\calO}
  \equiv \frac{\int_{\bbR^N} dx \, e^{-S(x)} \, \calO(x)}
  {\int_{\bbR^N} dx \, e^{-S(x)}},
\label{vev}
\end{align}
where $S(x) \in \bbC$ is a complex-valued action 
and $dx \equiv dx^1\wedge\cdots\wedge dx^N$ is the flat measure.
Since the Boltzmann weight $e^{-S(x)} / \int_{\bbR^N} dx \, e^{-S(x)}$ 
does not serve as a real and positive probability density, 
Monte Carlo methods based on importance sampling cannot be directly applied. 
A standard workaround is the so-called naive reweighting method, 
which defines a sampling measure using the real part of the action 
and expresses $\vev{\calO}$ as a ratio of reweighted averages:
\begin{align}
  \vev{ \mathcal{O} }
  = \frac{
    \int_{\bbR^N} dx \, e^{-\ReS(x)} \,  e^{-i \ImS(x)}\,\calO(x) 
    \,/\,\int_{\bbR^N} dx \, e^{-\ReS(x)} 
  }{ 
    \int_{\bbR^N} dx \, e^{-\ReS(x)} \, e^{-i \ImS(x)}
    \,/\,\int_{\bbR^N} dx \, e^{-\ReS(x)} 
  }
  \equiv \frac{ \vev{e^{-i \ImS(x)}\,\calO(x) }_{\mathrm{rewt}} }
         { \vev{e^{-i \ImS(x)} }_{\mathrm{rewt}} }.
\label{observable_naive_reweighting}
\end{align}
However, for systems with a large number of degrees of freedom ($N \gg 1$),
both the numerator and the denominator involve highly oscillatory integrals, 
giving exponentially small signals of order $e^{-O(N)}$. 
This renders numerical evaluation via Markov chain Monte Carlo methods impractical, 
because the sample size $N_{\text{conf}}$ must be exponentially large 
to make the statistical errors [of order $O(1/\sqrt{N_{\text{conf}}})$] 
relatively smaller than the signals. 

\begin{figure}[t]
  \centering
  \includegraphics[width=90mm]{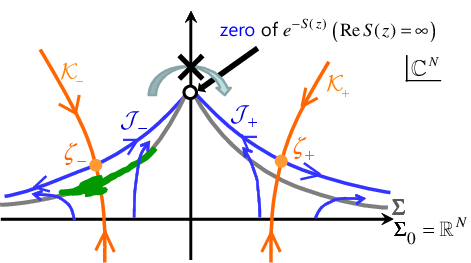}
  \caption{
    Lefschetz thimble method and its ergodicity problem 
    (figure adapted from Ref.~\cite{Fukuma:2023eru}). 
    The original integration surface $\Sigma_0 = \bbR^N$ 
    is deformed into $\Sigma$ within $\bbC^N$. 
    $\calJ_{\pm}$ ($\calK_{\pm}$) are 
    Lefschetz thimbles (anti-thimbles) 
    associated with critical points $\zeta_\pm$. 
    A Monte Carlo walker in the vicinity of one thimble $\calJ_-$
    cannot move to that of another thimble $\calJ_+$ 
    due to the infinitely high potential barrier at zeros of $e^{-S(z)}$.
  }
  \label{fig:ergodicity_problem}
\end{figure}

In the Lefschetz thimble method, 
the original integration surface $\Sigma_0 = \bbR^N$ is deformed 
into a submanifold $\Sigma$ within the complexified space $\bbC^N$, 
so that the oscillatory behavior of the integrand is alleviated on $\Sigma$
(see Fig.~\ref{fig:ergodicity_problem}).
When $e^{-S(z)}$ and $e^{-S(z)} \calO(z)$ are holomorphic over $\bbC^N$ 
(which is typically the case in physical models), 
the value of the integral remains unchanged 
under the deformation that fixes the boundary, 
as guaranteed by Cauchy's theorem: 
\begin{align}
  \langle \calO \rangle
  = \frac{\int_{\Sigma} dz \, e^{-S(z)} \, \calO(z)}
  {\int_{\Sigma} dz \, e^{-S(z)}} ,
\label{observable_GT_HMC}
\end{align}
where $dz \equiv dz^1\wedge\cdots\wedge dz^N$ is the holomorphic $N$-form. 
Such a deformation can be achieved 
by integrating the anti-holomorphic gradient flow:
\begin{align}
  \dot{z} &= \overline{\partial S(z)}
  \with
  z|_{t=0} = x.
\label{flow_config}
\end{align}
Here, $\dot{z} = \partial z / \partial t$, 
$t$ is the deformation parameter (referred to as the flow time), 
and $x$ is an initial configuration 
on the original integration surface $\Sigma_0$, 
with which a point on the flow is uniquely specified as $z = z(t,x)$. 
Due to the (in)equality 
\begin{align}
  [S(z)]^\centerdot = \partial S(z)\cdot \dot{z}
  = |\partial S(z)|^2 \geq 0,
\end{align}
we see that the real part $\ReS(z)$ always increases under the flow 
except at critical points 
[where the gradient $\partial S(z) = (\partial_i S(z))$ vanishes] 
while the imaginary part $\ImS(z)$ is kept constant under the flow. 

As the flow time $t$ increases, 
the deformed surface 
$\Sigma_t = \{ z = z(t,x) \mid x \in \Sigma_0 \}$ 
approaches a union of Lefschetz thimbles. 
Here, the Lefschetz thimble $\calJ$ associated with a critical point $\zeta$ 
is defined as the set of points flowing out of $\zeta$, 
on which $\ImS(z)$ is constant, $\ImS(z) = \ImS(\zeta)$ ($z \in \calJ$) 
(see Fig.~\ref{fig:ergodicity_problem}).
We thus expect that 
the oscillatory behavior of the integrand is significantly suppressed 
when the flow time $t$ becomes sufficiently large. 
However, the zeros of $e^{-S(z)}$ separate the deformed surface $\Sigma$, 
and thus Monte Carlo sampling on $\Sigma$ has an ergodicity problem.%
\footnote{
  The interesting proposal of the \emph{generalized thimble method} 
  \cite{Alexandru:2017lqr} 
  is to choose a deformed integration surface at an intermediate flow time 
  so as to alleviate 
  both the sign problem and the ergodicity problem simultaneously. 
  However, a detailed analysis \cite{Fukuma:2019wbv} shows that,
  in most cases, 
  the sign problem begins to relax 
  only after the deformed surface touches the zeros of $e^{-S(z)}$. 
} 
The WV-HMC method addresses this problem and is introduced 
as follows \cite{Fukuma:2020fez}. 

\begin{figure}[t]
  \centering
  \includegraphics[width=90mm]{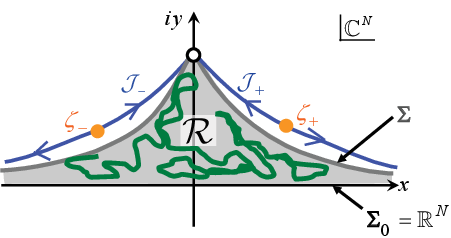}
  \caption{
   Worldvolume $\calR$
   (figure adapted from Ref.~\cite{Fukuma:2023eru}). 
   $\zeta_\pm$ are critical points, and $\calJ_\pm$ are 
   their associated Lefschetz thimbles. 
   The white circle stands for zeros of $e^{-S(z)}$ 
   which behave as infinitely high potential barriers 
   for a Monte Carlo walker. 
  }
  \label{fig:worldvolume}
\end{figure}

Since neither the numerator nor the denominator in Eq.~\eqref{observable_GT_HMC} 
depends on $t$,
we can take averages over flow time $t$ 
with an arbitrary common weight $e^{-W(t)}$:
\begin{align}
  \langle \calO \rangle
  = \frac{\int dt \, e^{-W(t)} \int_{\Sigma_t} dz \, e^{-S(z)}\,\calO(z)}
  {\int dt \, e^{-W(t)} \int_{\Sigma_t} dz \, e^{-S(z)}},
\label{observable_WV_HMC}
\end{align}
which can be viewed 
as a ratio of path integrals over the \emph{worldvolume} $\calR$ 
(see Fig.~\ref{fig:worldvolume}) defined by 
\begin{align}
  \calR\equiv \bigcup_{t} \Sigma_t=\{z(t,x) \mid t\in\bbR,\,x\in\bbR^N\}. 
\end{align} 
Monte Carlo sampling over $\calR$ provides detours 
that connect regions originally separated by the potential barriers. 

\begin{figure}[t]
  \centering
  \includegraphics[width=70mm]{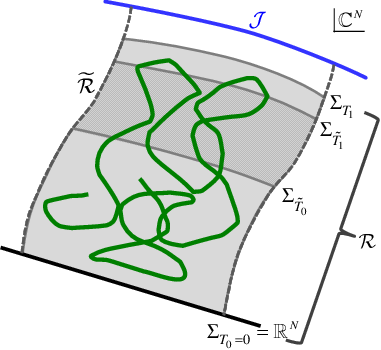}
  \caption{
    Worldvolume $\calR$ 
    (figure adapted from Ref.~\cite{Fukuma:2025gya}). 
    Measurements can be restricted to a subregion $\tilde{\calR}$ 
    to improve the signal-to-noise ratio 
    \cite{Fukuma:2020fez,Fukuma:2021aoo}.
  }
  \label{fig:md_flat}
\end{figure}

The extent of the worldvolume $\calR$ in the flow-time direction 
can be effectively restricted to a finite interval $[T_0,T_1]$ 
by adjusting the functional form of $W(t)$, 
which we choose as follows (see Fig.~\ref{fig:md_flat})
\cite{Fukuma:2023eru}:
\begin{align}
  W(t) = 
  \begin{cases}  
    -\,\gamma(t-T_0) + c_0\,\bigl(e^{(t-T_0)^2/2d_0^2} - 1\bigr) 
    &  \text{for } t < T_0
    \\
    -\,\gamma(t-T_0)                                             
    &  \text{for } T_0 \leq t \leq T_1
    \\
    -\,\gamma(t-T_0) + c_1\,\bigl(e^{(t-T_1)^2/2d_1^2} - 1\bigr) 
    &  \text{for } t > T_1 .            \\
  \end{cases}
  \label{Wt}
\end{align}
Here, $c_0, d_0$ and $c_1, d_1$ determine the heights and penetration depths 
of the potential walls placed at $t \sim T_0$ and $t \sim T_1$, respectively. 
The tilt parameter $\gamma$ introduces a constant driving force 
to prevent configurations from accumulating at small flow times.
These parameters are tuned 
to achieve an approximately uniform distribution of configurations 
along flow times.
The lower cutoff $T_0$ is chosen sufficiently small 
to ensure ergodicity 
on surfaces $\Sigma_t$ at $t \sim T_0$,
while the upper cutoff $T_1$ is taken to be sufficiently large 
so that the oscillatory behavior is well suppressed at $t \sim T_1$.
The latter condition is verified 
by computing the average reweighting factor $\vev{ \calF(z) }_{\Sigma_t}$ 
at various flow times $t$ using GT-HMC,%
\footnote{ 
  The HMC algorithm 
  using the RATTLE integrator \cite{Andersen:1983,Leimkuhler:1994} 
  was first introduced to the Lefschetz thimble method 
  in the seminal paper by the Komaba group \cite{Fujii:2013sra}, 
  where HMC updates are performed 
  directly on a single dominant Lefschetz thimble. 
  Its generalization to a deformed surface $\Sigma = \Sigma_t$ at fixed $t$ 
  was developed in Refs.~\cite{Alexandru:2019,Fukuma:2019uot}.
  The latter approach can be viewed 
  as an HMC version of the generalized thimble method \cite{Alexandru:2015sua}, 
  and is thus referred to 
  as the \emph{generalized thimble Hybrid Monte Carlo} (GT-HMC) method 
  in this paper. 
} 
which performs HMC updates on a single $\Sigma_t$ 
\cite{Alexandru:2019,Fukuma:2019uot}.
Configurations for measurement can be restricted further, if necessary, 
within a subinterval $[\tilde{T}_0,\tilde{T}_1]$, 
corresponding to the region $\tilde{\calR}$ shown in Fig.~\ref{fig:md_flat}, 
to exclude the lower region contaminated with the sign problem 
and the upper region that may by hard to sample correctly 
due to the complicated geometry at large flow times 
\cite{Fukuma:2020fez,Fukuma:2021aoo}. 
%

The expression \eqref{observable_WV_HMC} 
can be rewritten as a ratio of reweighted averages over $\calR$ 
as \cite{Fukuma:2020fez} 
\begin{align}
  \vev{ \calO }
  &= \frac{\langle \calF(z) \, \calO(z) \rangle_\calR}
  {\langle \calF(z) \rangle_\calR},
\label{vev_R_ratio}
\\
  \vev{ g(z) }_\calR 
  &= \frac{\int_\calR\, |dz|_\calR\,e^{-V(z)}\,g(z)}
  {\int_\calR\, |dz|_\calR\,e^{-V(z)}}.
\label{vev_R}
\end{align}
Here, $V(z)$ is the potential defined by 
\begin{align}
  V(z) = \ReS(z) + W(t(z)),
\end{align}
$|dz|_\calR$ is the volume element on $\calR$,
and $\calF(z)$ is the associated reweighting factor,%
\footnote{ 
  $E = (E^i_a \equiv \partial z^i/\partial x^a)$ is the Jacobian matrix. 
  In this paper, 
  $T_z \calS$ ($N_z \calS$) denotes the tangent (normal) space at $z$ 
  to a submanifold $\calS \subset \bbC^N$. 
  The flow vector $\xi \equiv \overline{\partial S}$ 
  is decomposed into its tangential and normal components, 
  $\xi = \xi_v + \xi_n$ 
  $(\xi_v \in T_z \Sigma_t,\,\,
  \xi_n \in N_z \Sigma_t$), 
  from which the lapse function $\alpha$ is defined as 
  $\alpha = \sqrt{\xi_n^\dagger \xi_n}$. 
  In GT-HMC \cite{Alexandru:2019,Fukuma:2019uot}, 
  the geometry in the flow-time direction is irrelevant, 
  so that the reweighting factor does not include the factor $\alpha^{-1}$ 
  and is given by a pure phase factor, 
  $(\det E / |\det E|)\,e^{-i\,\ImS} \equiv e^{i \theta}$.  
  \label{fn:reweighting_factor}
} 
\begin{align}
  \calF(z) 
  \equiv \frac{dt\,dz}{|dz|_\calR}\,e^{-i \ImS(z)}
  = \alpha^{-1}\,\frac{\det E}{|\det E |}\,e^{-i\,\ImS(z)}. 
\end{align}
The reweighted average on $\calR$ [Eq.~\eqref{vev_R}]
can be written as a phase-space integral of the form 
\cite{Fukuma:2020fez,Fukuma:2023eru,Fukuma:2025gya} 
\begin{align}
  \vev{ g(z) }_\calR 
  = \frac{\int_{T\calR}\, d\Omega_\calR\,e^{-H(z,\pi)}\,g(z)}
  {\int_{T\calR}\, d\Omega_\calR\,e^{-H(z,\pi)}}. 
\end{align}
Here, $\pi \in T_z\calR$ is the momentum,%
\footnote{ 
  More precisely, $\pi$ is the velocity.
} 
and $T\calR$ is the tangent bundle over $\calR$, 
\begin{align}
  T\calR = \{ (z,\pi) \mid z \in \calR, \,\pi \in T_z \calR \}, 
\end{align}
which carries a natural symplectic structure 
with symplectic form 
\begin{align}
  \omega_\calR \equiv \re \bigl( \overline{d\pi^i} \wedge dz^i \bigr). 
\end{align}
The symplectic volume form $d\Omega_\calR$ is given by 
\begin{align}
  d\Omega_\calR = \frac{\omega_\calR^{N+1}}{(N+1)!}, 
\end{align}
and the Hamiltonian $H(z,\pi)$ takes the form 
\begin{align}
  H(z,\pi) = \frac{1}{2}\,\pi^\dagger \pi + V(z).
\end{align}

\begin{figure}[t]
  \centering
  \includegraphics[width=70mm]{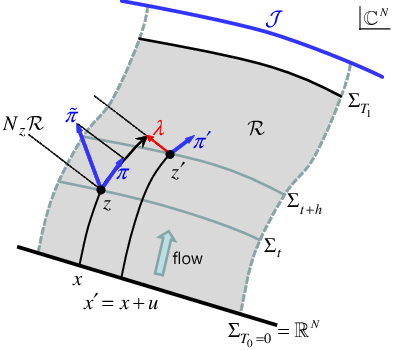}
  \caption{
    MD step on the worldvolume 
    (figure adapted from Ref.~\cite{Fukuma:2023eru}). 
  }
  \label{fig:wv_rattle}
\end{figure}

Molecular dynamics (MD) on $T\calR$ (see Fig.~\ref{fig:wv_rattle}) 
is generated by a RATTLE-type integrator \cite{Andersen:1983,Leimkuhler:1994}
of the following form 
\cite{Fukuma:2020fez} (see also Refs.~\cite{Fukuma:2023eru,Fukuma:2025gya}):
\begin{align}
  \pi_{1/2} &= \pi - \Delta s\,\overline{\partial V(z)} - \Delta s\,\lambda,
\label{wv_rattle1}
\\
  z' &= z + \Delta s\,\pi_{1/2},
\label{wv_rattle2}
\\
  \pi' &= \pi_{1/2} - \Delta s\,\overline{\partial V(z')} - \Delta s\,\lambda'.
\label{wv_rattle3}
\end{align}
Here, $\Delta s$ denotes the step size, 
and the force $-\overline{\partial V(z)}$ can be set to the following form
\cite{Fukuma:2020fez,Fukuma:2023eru}: 
\begin{align}
  -\overline{\partial V(z)} 
  = -\frac{1}{2}\,\Bigl[
    \xi + \frac{W'(t)}{\xi_n^\dag\, \xi_n}\,\xi_n
    \Bigr].
\label{force}
\end{align} 
The Lagrange multipliers $\lambda\in N_z\calR$ and $\lambda'\in N_{z'}\calR$ 
are determined so that $z'\in \calR$ and $\pi'\in T_{z'}\calR$, respectively. 
The former condition ($z' \in \calR$) is equivalent to finding a triplet 
$\{ h \in \bbR,\,u \in T_x\Sigma_0,\,\lambda \in N_z \calR \}$ 
that satisfies the following relation 
for given $z = z(t,x) \in \calR$ and $\pi \in T_z \calR$: 
\begin{align}
  z(t+h, x+u) 
  = z + \Delta s\,[\pi - \Delta s\,\overline{\partial V(z)} - \Delta s\,\lambda],
\end{align}
whose solution can be obtained via simplified Newton iteration 
(see Ref.~\cite{Fukuma:2023eru} for details). 
The latter condition ($\pi' \in T_{z'}\calR$) is realized 
by projecting the vector 
$\pi_{1/2} - \Delta s\,\overline{\partial V(z')}$ 
onto the tangent space $T_{z'}\calR$ at $z'$. 
The transformation $(z,\pi) \to (z',\pi')$ 
[Eqs.~\eqref{wv_rattle1}--\eqref{wv_rattle3}] 
is exactly reversible 
and symplectic ($\omega'_\calR = \omega_\calR$), 
and thus volume-preserving ($d\Omega'_\calR = d\Omega_\calR$). 
Moreover, the energy is conserved up to second order in $\Delta s$, 
i.e., $H(z',\pi') = H(z,\pi) + O(\Delta s^3)$. 
%

\begin{figure}[t]
  \centering
  \includegraphics[width=70mm]{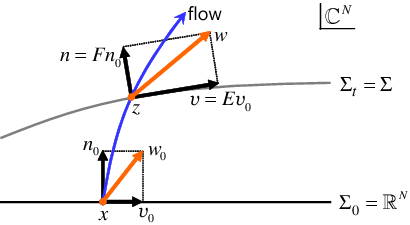}
  \caption{
    Deformation of the integration surface 
    (figure adapted from Ref.~\cite{Fukuma:2023eru}). 
    A tangent vector $v_0 \in T_x \Sigma_0$ and 
    a normal vector $n_0 \in N_x \Sigma_0$ at $x$ 
    are lifted to 
    $v \in T_z \Sigma$ and $n \in N_z \Sigma$ at $z$, respectively. 
  }
  \label{fig:flow}
\end{figure}

The computational cost of both GT-HMC and WV-HMC 
is dominated by the task of integrating the vector flow equations 
for a tangent vector $v \in T_z \Sigma_t$ 
and a normal vector $n \in N_z \Sigma_t$ 
(see Fig.~\ref{fig:flow}) \cite{Alexandru:2017lqr}, 
\begin{align}
  \dot{v} 
  &= \overline{H(z) v}
  \quad\text{with}\quad
  v|_{t=0} = v_0,
\label{flow_tangent}
\\
  \dot{n} 
  &= -\overline{H(z) n}
  \quad\text{with}\quad
  n|_{t=0} = n_0,
\label{flow_normal}
\end{align}
where $H_{ij}(z) \equiv \partial_i \partial_j S(z)$ is the Hessian matrix. 
Provided that $H$ is a sparse matrix 
and the convergence rate of iterations in the RATTLE update 
depends only weakly on the system size, 
the computational cost is expected to be $O(N)$, 
as is the case in complex scalar field theory at finite density 
\cite{Namekawa:2024ert}. 
%

In cases where the Boltzmann weight includes the fermion determinant 
(as in the DQMC computation of the Hubbard model), 
we need to invert fermion matrices. 
Two approaches are possible \cite{Fukuma:2025esu}. 
One is to use direct solvers for the inversion, 
which results in a computational cost of $O(N^3)$. 
The other is to employ pseudofermions and iterative solvers. 
This reduces the cost to $O(N^2)$, 
but requires careful tuning of parameters 
to justify the use of pseudofermions 
and to ensure proper convergence of the iterative solver \cite{Fukuma:2025esu}. 
In this paper, 
we focus exclusively on the first approach. 
A detailed discussion of the second approach 
will be presented in a separate publication.

\section{Path-integral representation of the Hubbard model}
\label{sec:path_integral}

In this section, 
we rewrite the grand canonical partition function of the Hubbard model 
in a path-integral form which is suitable for WV-HMC simulations.

\subsection{The Hubbard model}
\label{sec:model}

The Hubbard model on a $d$-dimensional spatial lattice 
is defined by the following Hamiltonian, 
which includes the chemical potential term: 
\begin{align}
  \hat H_\mu^{\mathrm{org}}
  &= \hat H - \mu \hat N
\nonumber
\\
  &\equiv
  - \sum_{\bx, \by}  \sum_{\sigma=\uparrow,\downarrow} 
    t_{\bx \by}\, c_{\bx, \sigma}^\dag c_{\by, \sigma}
  + U      \sum_{\bx} n_{\bx, \uparrow} n_{\bx, \downarrow} 
  - \mu    \sum_{\bx} (n_{\bx, \uparrow} + n_{\bx, \downarrow}) .
\label{hamiltonian_org}
\end{align}
Here,  $c_{\bx, \sigma}$ ($c_{\bx, \sigma}^\dag$)  
denotes the annihilation (creation) operator of an electron 
with spin $\sigma$ ($=\uparrow,\,\downarrow$) at site $\bx=(x_i)$ ($i=1,\ldots,d$), 
and $n_{\bx, \sigma} \equiv c_{\bx, \sigma}^\dag c_{\bx, \sigma}$. 
The hopping matrix $t_{\bx \by}$ is assumed to take the form%
\footnote{ 
  The same symbol $t$ is used 
  for both the flow time and the hopping matrix (or hopping amplitude). 
  The intended meaning should be clear from the context. 
} 
\begin{align}
  t_{\bx \by} = 
  \begin{cases}
     t \ (>0) & \text{if } \bx \text{ and } \by \text{ are nearest neighbors}, \\
     0        & \text{otherwise}.
  \end{cases}
\label{hopping}
\end{align} 
The parameters $U$ and $\mu$ denote 
the on-site interaction strength and the chemical potential, respectively. 
The total number operator is given by  
$\hat N = \sum_\bx \sum_\sigma n_{\bx,\sigma}$.
We assume that the model is defined on a periodic, bipartite square lattice 
with linear extent $L_s$, 
so that the spatial volume is given by $V_d \equiv L_s^d$. 
To ensure that 
the bosonized action (introduced below) 
is manifestly real-valued at half filling, 
we perform a particle-hole transformation on the spin-down component as 
\begin{align}
  a_\bx \equiv c_{\bx\uparrow},\quad
  b_\bx \equiv (-1)^\bx c^\dag_{\bx\downarrow},
\end{align}
where $(-1)^\bx\equiv (-1)^{\sum_i x_i}$ is the site parity. 
Up to an additive constant, 
the Hamiltonian \eqref{hamiltonian_org} is then rewritten as
\begin{align}
  \hat H_\mu 
  \equiv \hat H_\mu^{\mathrm{org}} - \mu V_d
  = - \sum_{\bx, \by} 
  t_{\bx \by}\, ( a_\bx^\dag a_\by + b_\bx^\dag b_\by )
  + \frac{U}{2}\, \sum_{\bx} ( n^a_\bx - n^b_\bx )^2
  - \tilde \mu\, \sum_{\bx} ( n^a_\bx - n^b_\bx ) ,
\label{hamiltonian}
\end{align}
where $n^a_\bx\ \equiv a_\bx^\dag a_\bx$ and $n^b_\bx\ \equiv b_\bx^\dag b_\bx$, 
and 
\begin{align}
  \tilde \mu \equiv \mu - \frac{U}{2}.
\end{align}
Note that 
under the exchange of $a$ and $b$, 
the first two terms in Eq.~\eqref{hamiltonian} remain invariant, 
whereas the last term changes sign. 
This implies that 
the continuum grand canonical partition function 
$\tr e^{-\beta \hat{H}_\mu}$ 
is an even function of $\tilde\mu$. 
Thus, we have $\vev{ n^a_\bx - n^b_\bx } = 0$ at $\tilde\mu = 0$, 
which means that 
the point $\tilde\mu = 0$ (i.e., $\mu = U/2$) 
corresponds to the half-filling state, 
$\vev{ n_{\bx,\uparrow} + n_{\bx,\downarrow} } = 1$.%
\footnote{ 
  This follows from $n_{\bx,\uparrow} + n_{\bx,\downarrow} = n^a_\bx - n^b_\bx + 1$.
} 

Following Ref.~\cite{Beyl:2017kwp}, 
we introduce a redundant parameter $\alpha$ $(0\leq\alpha\leq 1)$ as%
\footnote{ 
  This equality directly follows from the identity 
  $ (n^a_\bx + n^b_\bx - 1)^2 = -(n^a_\bx - n^b_\bx)^2 + 1$,  
  which holds because $(n^{a/b}_\bx)^2 = n^{a/b}_\bx$ 
  \cite{Beyl:2017kwp}. 
} 
\begin{align}
  (n^a_\bx - n^b_\bx)^2 
  = \alpha (n^a_\bx - n^b_\bx)^2
  - (1-\alpha) (n^a_\bx + n^b_\bx -1)^2 + 1 - \alpha,
\label{alpha}
\end{align}
and rewrite the Hamiltonian in the form 
\begin{align}
  \hat H_\mu \equiv \hat H_\mu^{(1)} + \hat H_\mu^{(2)} 
\end{align}
with $\hat H_\mu^{(1)}$ and $\hat H_\mu^{(2)}$ 
denoting the one-body and two-body parts of the Hamiltonian, respectively: 
\begin{align}
  \hat H_\mu^{(1)} 
  &\equiv
    -\sum_{\bx,\by}\, (a_\bx^\dag, b_\bx^\dag)
    \begin{pmatrix}
      t_{\bx\by} + \tilde\mu\,\delta_{\bx\by} & 0 \\
      0 & t_{\bx\by} - \tilde\mu\,\delta_{\bx\by}
    \end{pmatrix}
    \begin{pmatrix}
      a_\by \\
      b_\by
    \end{pmatrix}
  \equiv c^\dag K c,
\\
  \hat H_\mu^{(2)} 
  &\equiv
    \frac{U}{2}\,\sum_\bx\,\bigl[
        \alpha (n^a_\bx - n^b_\bx)^2
      - (1-\alpha) (n^a_\bx + n^b_\bx -1)^2 + 1 - \alpha
    \bigr].
\end{align}
Here, we have introduced a doublet operator $c$ and its conjugate $c^\dag$ as 
\begin{align}
  c = 
  \begin{pmatrix}
    a = (a_\bx) \\ 
    b = (b_\bx)
  \end{pmatrix},
  \quad
  c^\dag = 
  \bigl( 
    a^\dag = (a_\bx^\dag),\; b^\dag = (b_\bx^\dag)
  \bigr) .
\end{align}
We divide the inverse temperature $\beta$ into $N_t$ time slices 
such that $\beta = N_t\,\epsilon$, 
and employ a symmetric imaginary-time evolution operator  
\begin{align}
  \hat T \equiv e^{-(\epsilon/2) \hat H_\mu^{(1)}}\,
    e^{-\epsilon \hat H_\mu^{(2)}}\, e^{-(\epsilon/2) \hat H_\mu^{(1)}} ,
\end{align}
which approximates the continuum evolution operator 
up to second order in $\epsilon$, 
$\hat T = e^{-\epsilon \hat H_\mu} + O(\epsilon^3)$. 
We then define the grand canonical partition function on the lattice 
as 
\begin{align}
  Z \equiv \tr\,\hat T^{N_t}
  ~ \bigl[\,
  = \tr e^{-\beta \hat H_\mu} + O(\epsilon^2) 
  = e^{\beta\mu V_d}\,\tr e^{-\beta \hat H_\mu^{\mathrm{org}}} + O(\epsilon^2)
  \,\bigr] .
\label{partition_function}
\end{align}

To rewrite $Z$ in a path-integral form, 
we introduce a set of Grassmann variables as 
\begin{align}
  \psi = 
  \begin{pmatrix}
    \psi_a = (\psi_{a,\bx}) \\ 
    \psi_b = (\psi_{b,\bx})
  \end{pmatrix},
  \quad
  \psi^\dag = 
  \bigl( 
  \psi_a^\dag = (\psi_{a,\bx}^\dag),\; \psi_b^\dag = (\psi_{b,\bx}^\dag)
  \bigr) ,
\end{align}
and define the coherent state of $c = (a,b)^T$ as 
\begin{align}
  \ket{\psi} \equiv e^{c^\dag \psi}\,\ket{0}
  &= e^{a^\dag \psi_a + b^\dag \psi_b }
      \,\ket{0},
\\
  \bra{\psi^\dag} \equiv \bra{0}\,e^{\psi^\dag c} 
  &= \bra{0}\,e^{ \psi_a^\dag a + \psi_b^\dag b }.
\end{align}
These states satisfy the following relations: 
\begin{align}
  \braket{\psi^\dag}{\psi'}
    &= e^{\psi^\dag \psi'} ,
\label{coherent1}
\\
  1 &= \int d\psi^\dag d\psi\, e^{-\psi^\dag \psi}
      \ket{\psi} \bra{\psi^\dag},
\label{coherent2}
\\
  \tr \calO 
    &= \int d\psi^\dag d\psi\, e^{-\psi^\dag \psi}\,
    \vev{ \psi^\dag\,|\, \calO \,| -\!\!\psi }
    \quad\text{(for bosonic operators $\calO$)},
\label{coherent3}
\end{align}
where the integration measure is defined by 
\begin{align}
  d\psi^\dag d\psi \equiv \prod_\bx d\psi_{a,\bx}^\dag d\psi_{a,\bx} 
  \prod_\bx d\psi_{b,\bx}^\dag d\psi_{b,\bx}. 
\end{align}
By repeatedly inserting the identity operator of the form \eqref{coherent2}, 
we obtain the Grassmann path-integral representation of $Z$: 
\begin{align}
  Z &= \int_{\mathrm{ABC}} 
  \Bigl(\prod_{\ell=1}^{N_t} d\psi_\ell^\dag d\psi_\ell \Bigr)\,
  e^{-\sum_\ell \psi_\ell^\dag \psi_\ell}
  \prod_{\ell=1}^{N_t}\,\bra{\psi_\ell^\dag}\, \hat T \,\ket{\psi_{\ell-1}},
\label{partition_function_psi}
\end{align}
where ``ABC'' denotes the anti-periodic boundary condition, 
$\psi_0 \equiv -\psi_{N_t}$.

\subsection{Bosonization}
\label{sec:bosonization}

Using the identity 
$e^{c^\dagger X c}\,\ket{\psi} = \ket{e^X \psi}$ 
for a $(2 V_d) \times (2 V_d)$  matrix $X$ acting on the doublet field, 
the matrix element of the transfer matrix can be written as 
\begin{align}
  \bra{\psi_\ell^\dag}\, \hat T \,\ket{\psi_{\ell-1}}
  &= \bra{\psi_\ell^\dag}\, 
    e^{-(\epsilon/2)\,c^\dag K c}\,
    e^{-\epsilon \hat H_\mu^{(2)}}\,
    e^{-(\epsilon/2)\,c^\dag K c}\,
    \ket{\psi_{\ell-1}}
\nonumber
\\
  &= \bra{ (e^{-(\epsilon/2)K}\psi_\ell)^\dag}\,
    e^{-\epsilon \hat H_\mu^{(2)}} 
    \ket{e^{-(\epsilon/2)K}\psi_{\ell-1}}. 
\label{transfer_mat1}
\end{align}
The operator $e^{-\epsilon \hat H_\mu^{(2)}}$ 
is diagonal in the site index $\bx$ 
and can be expressed as a Gaussian integral 
over two auxiliary (Hubbard-Stratonovich) fields \cite{Beyl:2017kwp}: 
\begin{align}
  &e^{-(\alpha\epsilon U/2)\,(n^a_\bx - n^b_\bx)^2 
      + ((1-\alpha)\epsilon U/2)\,(n^a_\bx + n^b_\bx - 1)^2 
      - (1-\alpha)\,\epsilon U/2}
\nonumber
\\
  &= \int dA_\bx dB_\bx\, e^{-(1/2)(A_\bx^2 + B_\bx^2)}\,
  e^{ [ i c_0 A_\bx + c_1 B_\bx - c_1^2] \, n^a_\bx} \,
  e^{ [-i c_0 A_\bx + c_1 B_\bx - c_1^2] \, n^b_\bx} 
\end{align}
with 
\begin{align}
  c_0 \equiv \sqrt{\alpha\epsilon U}, \quad
  c_1 \equiv \sqrt{(1-\alpha)\epsilon U} .
\end{align}
This leads to the identity 
\begin{align}
  &\bra{\psi^\dag}\,e^{-\epsilon \hat H_\mu^{(2)}}\,\ket{\psi'}
  = \int dA\, dB\, 
    e^{-(1/2)\sum_\bx (A_\bx^2 + B_\bx^2)}
    \exp\bigl[
      \psi_a^\dag\, 
      e^{i c_0 A + c_1 B - c_1^2}
      \psi'_a
      +
      \psi_b^\dag\, 
      e^{-i c_0 A + c_1 B - c_1^2}
      \psi'_b
    \bigr],
\label{transfer_mat2}
\end{align}
where $A = (A_\bx\delta_{\bx\by})$, $B = (B_\bx\delta_{\bx\by})$, 
and $dA\, dB = \prod_\bx dA_\bx dB_\bx$. 
Substituting Eqs.~\eqref{transfer_mat1} and \eqref{transfer_mat2} 
into Eq.~\eqref{partition_function_psi}, 
the partition function becomes 
\begin{align}
  Z 
  &= 
  \int \Bigl( \prod_\ell dA_\ell dB_\ell \Bigr) \,
  e^{-(1/2)\sum_{\ell,\bx}\,(A_{\ell,\bx}^2 + B_{\ell,\bx}^2)}\,
  \int_{\mathrm{ABC}} 
  \Bigl( \prod_\ell  
     d\psi_{\ell}^\dag d\psi_{\ell}
  \Bigr)\,
  e^{ -\sum_\ell \psi_{\ell}^\dag \psi_{\ell} } \,
\nonumber
\\
  &\quad \times 
  \exp\sum_\ell \bigl[
    (e^{-(\epsilon/2) K} \psi_\ell^\dag)_a\,
    e^{i c_0 A_\ell + c_1 B_\ell - c_1^2}\,
    (e^{-(\epsilon/2) K}\psi_{\ell-1})_a
  \bigr]\,
\nonumber
\\
  &\quad \times 
  \exp\sum_\ell \bigl[
    (e^{-(\epsilon/2) K}\psi_\ell^\dag)_b\,
    e^{-i c_0 A_\ell + c_1 B_\ell - c_1^2}\,
    (e^{-(\epsilon/2) K}\psi_{\ell-1})_b
  \bigr]\,
\nonumber
\\
  &= \int \Bigl( \prod_{\ell,\bx} dA_{\ell,\bx} dB_{\ell,\bx} \Bigr) \,
  e^{-(1/2)\sum_{\ell,\bx}\,(A_{\ell,\bx}^2 + B_{\ell,\bx}^2)}\,
\nonumber
\\
  &\quad \times 
  \int_{\mathrm{ABC}} 
  \Bigl(
    \prod_{\ell,\bx} 
    d(\psi_a^\dag)_{\ell,\bx} d(\psi_a)_{\ell,\bx}\,
    \prod_{\ell,\bx} 
    d(\psi_b^\dag)_{\ell,\bx} d(\psi_b)_{\ell,\bx}
  \Bigr)  
  e^{-\sum_{\ell,\bx}\bigl[
    (\psi_a^\dag)_{\ell,\bx}\,(\tilde{D}_a\psi_a)_{\ell,\bx}
    + (\psi_b^\dag)_{\ell,\bx}\,(\tilde{D}_b\psi_b)_{\ell,\bx}
  \bigr]} ,
\end{align}
where 
\begin{align}
  (\tilde{D}_{a/b}\,\psi_{a/b})_{\ell}
  = (\psi_{a/b})_{\ell} 
    - e^{(\epsilon /2)\, t}\,
      e^{\pm (\epsilon\tilde\mu + i c_0 A_\ell) + c_1 B_\ell - c_1^2}\,
      (e^{(\epsilon /2)\, t}\,\psi_{a/b})_{\ell-1}, 
\end{align}
and we have rewritten $(\psi_\ell)_{a/b,\bx}$ and $(\psi_\ell^\dagger)_{a/b,\bx}$ 
as $(\psi_{a/b})_{\ell,\bx}$ and $(\psi_{a/b}^\dagger)_{\ell,\bx}$, 
respectively.

We now introduce a $(d+1)$-dimensional spacetime lattice 
of volume $V_{d+1} \equiv N_t \times  V_d = N_t \times L_s^d$  
with coordinates denoted by $x = (\ell,\bx)$. 
We regard $A \equiv (A_x = A_{\ell,\bx})$ 
and $B \equiv (B_x = B_{\ell,\bx})$ 
as fields defined on this spacetime lattice 
(reusing the symbols $A$ and $B$). 
We also define $V_{d+1} \times V_{d+1}$ matrices 
indexed by $x=(\ell,\bx)$ and $y=(m,\by)$ 
as follows (reusing the symbol $t$): 
\begin{align}
  t  &= (t_{xy}) 
  \quad\text{with}\quad 
  t_{x y} = (\delta_{\ell m}\,t_{\bx\by} ),
\\
  \Lambda_0 &= ((\Lambda_0)_{xy}) 
   \quad\text{with}\quad (\Lambda_0)_{xy} = 
   \left\{ \begin{array}{ll}
      \delta_{\ell+1,m}\,\delta_{\bx\by} & (\ell < N_t) \\
      - \delta_{1,m}\,\delta_{\bx\by}      & (\ell = N_t).
   \end{array}\right. 
\end{align} 
Then, the fermion operators $\tilde{D}_{a/b}$ can be written as 
\begin{align}
  \tilde{D}_{a/b} = 
    1 - e^{(\epsilon/2)\,t}\,
    h_{a/b}\,
    \Lambda_0^{-1}\,e^{(\epsilon/2)\,t}
  \equiv -e^{(\epsilon/2)\, t}\,D_{a/b}\,\Lambda_0^{-1}\,
  e^{(\epsilon/2)\,t}. 
\end{align}
Here,
$h_{a/b} = ( (h_{a/b})_x\,\delta_{x y} )$ are diagonal matrices 
defined by 
\begin{align}
  (h_{a/b})_x \equiv e^{\pm (\epsilon\tilde\mu + i c_0 A_x) + c_1 B_x - c_1^2} ,
\end{align}
and 
\begin{align}
  D_{a/b}(A,B) \equiv 
  h_{a/b}
  - e^{-\epsilon\,t}\,\Lambda_0 .
\end{align}
Finally, using the identities  
$\det e^{(\epsilon/2)\,t} = e^{(\epsilon/2) \,\tr t} = 1$ 
and $\det \Lambda_0 = 1$, 
we arrive at the bosonized form of the partition function:  
\begin{align}
  Z = \int dA\,dB\,e^{-S(A,B)}
    = \int dA\,dB\,e^{-(1/2)\sum_x (A_x^2 + B_x^2)}\,\det D_a(A,B)\,\det D_b(A,B), 
\label{path_int_AB}
\end{align} 
where the measure is now defined as $dA\, dB \equiv \prod_x dA_x\, dB_x$. 
The partition function $Z$ in the path-integral form 
remains an even function of $\tilde\mu$, 
because the change of variable $A \to -A$ renders 
$D_{a/b} \to D_{b/a} |_{\tilde\mu \to -\tilde\mu}$. 
Also, $Z$ is real-valued, 
because the same transformation changes 
$D_{a/b} \to \overline{D_{a/b}}$.

Note that at half filling $\tilde\mu = 0$, 
we have $D_b = \overline{D_a}$, 
and thus $\det D_a \det D_b = |\det D_a|^2$. 
This implies that 
the path integral is free from the sign problem at half filling. 
Furthermore, 
even when $\tilde\mu \neq 0$, 
the sign problem is expected to remain mild 
as long as $D_b \approx \overline{D_a}$. 
Since the auxiliary fields $A$ and $B$ fluctuate around zero and $c_1$, respectively,  
with variance of order $O(1)$, 
this approximate equality is expected to hold  
when $\epsilon\tilde\mu \ll c_1 = \sqrt{(1-\alpha)\epsilon U}$, 
i.e., when $\alpha \ll 1 - \epsilon\tilde\mu^2/U$. 

At $\alpha = 0$, 
the determinants $\det D_{a/b}$ are real on $\Sigma_0$ but can vanish. 
Consequently, $\Sigma_0$ becomes partitioned into disjoint regions 
defined by $\det D_a \det D_b \gtrless 0$. 
This partitioning creates ergodicity barriers on $\Sigma_0$, 
rendering the WV-HMC scheme invalid. 
While introducing a non-zero $\alpha$ alleviates these ergodicity issues, 
excessively large values of $\alpha$ exacerbate the sign problem 
and increase the computational cost. 
Thus, $\alpha$ must be carefully tuned within a specific window: 
it must be large enough to ensure ergodicity on $\Sigma_0$, 
yet sufficiently small to effectively suppress the sign problem 
\cite{Beyl:2017kwp}.%
\footnote{ 
  Another prescription for enhancing ergodicity 
  is to use a negative flow-time cutoff $T_0$ ($T_0 < 0$) \cite{Fukuma:2017fjq,Fukuma:2020fez}.
} 
We perform this tuning of $\alpha$ in Sect.~\ref{sec:results2}.

\section{Applying WV-HMC to the Hubbard model}
\label{sec:app}

We start from the following action 
involving two dynamical fields $A$ and $B$:
\begin{align}
  S(A,B) = \frac{1}{2}\,\sum_x (A_x^2 + B_x^2) 
  - \ln\det D_a(A,B) - \ln\det D_b(A,B) .
\end{align}
Taking into account that our evolution operator $\hat{T}$ 
approximates the continuum one $e^{-\epsilon \hat{H}_\mu}$ 
only up to second order in $\epsilon$, 
we approximate $D_{a/b}$ to the same order:%
\footnote{ 
  Note that $\Lambda_0$ should not be estimated 
  as $\Lambda_0 = 1 + O(\epsilon)$, 
  because this relation holds only for thermalized configurations. 
} 
\begin{align}
  D_{a/b} = h_{a/b} - \Lambda_0 + \epsilon\, t\, \Lambda_0
    - \frac{\epsilon^2}{2}\,t^2 \Lambda_0. 
\end{align}

\subsection{Flow equations}
\label{sec:wv1_flow_eqs}

Using the identities 
\begin{align}
  \frac{\partial (h_{a/b})_y}{\partial A_x} 
  &= \pm i c_0 (h_{a/b})_x\,\delta_{xy},
  &
  \frac{\partial (h_{a/b})_y}{\partial B_x} 
  &= c_1 (h_{a/b})_x\,\delta_{xy},
\\
  \frac{\partial (D_{a/b})_{yz}}{\partial A_x} 
  &= \pm i c_0 (h_{a/b})_x\,\delta_{xyz},
  &
  \frac{\partial (D_{a/b})_{yz}}{\partial B_x} 
  &= c_1 (h_{a/b})_x\,\delta_{xyz} 
  \quad
  (\delta_{xyz}\equiv \delta_{xy}\delta_{yz}) ,
\end{align}
and defining the combinations $v_{a/b}$ 
from a doublet field $v = (v^A_x,v^B_x)$ as
\begin{align}
  (v_{a/b})_x \equiv \pm i\, c_0 v^A_x + c_1 v^B_x,
\end{align}
we obtain the gradient $\partial S$ of the action 
and the Hessian operator $H$ acting on a doublet field $v=(v^A_x,v^B_x)$ 
as follows: 
\begin{align}
  (\partial S)^A_x 
  &\equiv \frac{\partial S}{\partial A_x}
  = A_x - i c_0\,\bigl[
      (D_a^{-1})_{xx} (h_a)_x - (D_b^{-1})_{xx} (h_b)_x
    \bigr],
\\
  (\partial S)^B_x 
  &\equiv \frac{\partial S}{\partial B_x}
  = B_x - c_1\,\bigl[
      (D_a^{-1})_{xx} (h_a)_x + (D_b^{-1})_{xx} (h_b)_x
    \bigr],
\\
  (Hv)^A_x 
  &= v^A_x - i c_0 \biggl[
    (h_a)_x \bigl[
      (D_a^{-1})_{xx}\, (v_a)_x
      - \sum_y (D_a^{-1})_{xy} (h_a)_y (v_a)_y (D_a^{-1})_{yx}
    \bigr]
\nonumber
\\
  &\quad
    - (h_b)_x \bigl[
        (D_b^{-1})_{xx}\, (v_b)_x
        - \sum_y (D_b^{-1})_{xy} (h_b)_y (v_b)_y (D_b^{-1})_{yx}
      \bigr]
    \biggr],
\\
  (Hv)^B_x 
  &= v^B_x - c_1 \biggl[
    (h_a)_x \bigl[
      (D_a^{-1})_{xx}\, (v_a)_x
      - \sum_y (D_a^{-1})_{xy} (h_a)_y (v_a)_y (D_a^{-1})_{yx}
    \bigr]
\nonumber
\\
  &\quad
    + (h_b)_x \bigl[
        (D_b^{-1})_{xx}\, (v_b)_x
        - \sum_y (D_b^{-1})_{xy} (h_b)_y (v_b)_y (D_b^{-1})_{yx}
      \bigr]
  \biggr] .
\end{align}
The flow equations for a configuration $(A_x,B_x)$, 
a tangent vector $v = (v^A_x, v^B_x)$, 
and a normal vector $n = (n^A_x, n^B_x)$ 
are then given by 
\begin{align}
  \dot{A}_x 
  &= \overline{(\partial S)^A_x},
  \quad
  \dot{B}_x = \overline{(\partial S)^B_x},
\\
  \dot{v}^A_x &= \overline{(Hv)^A_x},
  \quad
  \dot{v}^B_x = \overline{(Hv)^B_x},
\\
  \dot{n}^A_x &= - \overline{(Hn)^A_x},
  \quad
  \dot{n}^B_x = - \overline{(Hn)^B_x}.
\end{align}
Once the flow equations are obtained, 
we only need to follow the general framework presented in Sect.~\ref{sec:wv-hmc}.

\subsection{Observables}
\label{sec:wv1_observables}

When the Trotter number $N_t$ is held fixed, 
the parameters $\beta$ and $\beta\mu$ enter the action 
only through $\epsilon$ 
and $\epsilon\tilde\mu = \epsilon\mu - \epsilon U/2$, respectively. 
We define 
the number density operator $n$ 
and the energy density operator $e$ 
as follows:%
\footnote{ 
  The term $+1$ comes from the fact that 
  the Hamiltonian \eqref{hamiltonian} 
  differs from the original Hamiltonian \eqref{hamiltonian_org} 
  by an additive constant $-\mu V_d$ 
  [see Eq.~\eqref{hamiltonian}]. 
} 
\begin{align}
  n(A,B) 
  &\equiv 
  - \frac{1}{V_{d+1}}\,
    \frac{\partial S(A,B)}{\partial (\epsilon\mu)}\Bigr|_{\epsilon}
  + 1
  = - \frac{1}{V_{d+1}}\,
    \frac{\partial S(A,B)}{\partial (\epsilon\tilde\mu)}\Bigr|_{\epsilon}
  + 1 ,
\label{number_density}
\\
  e(A,B) 
  &\equiv 
  \frac{\partial S(A,B)}{\partial \epsilon}\Bigr|_{\epsilon\mu}
  = \frac{1}{V_{d+1}}\,\biggl[
  \frac{\partial S(A,B)}{\partial \epsilon}\Bigr|_{\epsilon\tilde\mu}
  - \frac{U}{2}\,
  \frac{\partial S(A,B)}{\partial (\epsilon\tilde\mu)}\Bigr|_{\epsilon}
  \biggr] .
\label{energy_density}
\end{align}%
Their expectation values can be estimated via the path integral, 
and are expected to agree with the continuum expectation values 
of $\hat N/V_d$ and $\hat H/V_d$ 
up to $O(\epsilon^2)$ corrections:
\begin{align}
  \langle n \rangle &\equiv 
  \frac{1}{V_{d+1}}\,\frac{\int (dA\,dB)\,e^{-S(A,B)}\,n(A,B)}
       {\int (dA\,dB) e^{-S(A,B)}}
  = \frac{1}{V_d}\,\frac{\tr\, e^{-\beta (\hat H - \mu \hat N)}\,\hat N }
                        {\tr\, e^{-\beta (\hat H - \mu \hat N)} }
    + O(\epsilon^2) ,
\label{scaling_n}
\\
 \langle e \rangle 
 &\equiv 
 \frac{1}{V_{d+1}}\,\frac{\int (dA\,dB)\,e^{-S(A,B)}\,e(A,B)}
 {\int (dA\,dB) e^{-S(A,B)}}
 = \frac{1}{V_d}\,\frac{\tr\, e^{-\beta (\hat H - \mu \hat N)}\,\hat H }
 {\tr\, e^{-\beta (\hat H - \mu \hat N)} }
 +  O(\epsilon^2). 
\label{scaling_e}
\end{align}%
Their explicit forms are given by%
\footnote{ 
  The following formulas will be useful for derivation,  
  which are actually exact to all orders in $\epsilon$ 
  [$A \equiv (A_x \delta_{xy})$, $B \equiv (B_x \delta_{xy})$ 
  and $t = (t_{xy})$]:
  \begin{align}
    \frac{\partial h_{a/b}}{\partial (\epsilon\tilde\mu)}\Bigr|_{\epsilon}
    &= \pm h_{a/b},
  \nonumber
  \\
    \frac{\partial h_{a/b}}{\partial \epsilon}\Bigr|_{\epsilon\tilde\mu}
    &= h_{a/b}\,\Bigl[
    \pm \frac{i}{2}\,\sqrt{\frac{\alpha U}{\epsilon}}\,A
    + \frac{1}{2}\,\sqrt{\frac{(1-\alpha) U}{\epsilon}}\,B
    - (1-\alpha) U  
    \Bigr], 
  \nonumber
  \\
    \frac{\partial}{\partial \epsilon}\,
    \Bigl( - e^{-\epsilon t}\,\Lambda_0^{-1} \Bigr)
    &= h_{a/b}\,t - D_{a/b}\,t,
  \nonumber
  \\
    \frac{\partial D_{a/b}}{\partial (\epsilon\tilde\mu)}\Bigr|_{\epsilon}
    &= \pm h_{a/b},
  \nonumber
  \\
    \frac{\partial D_{a/b}}{\partial \epsilon}\Bigr|_{\epsilon\tilde\mu}
    &= h_{a/b}\,\Bigl[
    \pm \frac{i}{2}\,\sqrt{\frac{\alpha U}{\epsilon}}\,A
    + \frac{1}{2}\,\sqrt{\frac{(1-\alpha) U}{\epsilon}}\,B
    - (1-\alpha) U + t 
    \Bigr] 
    - D_{a/b}\, t.
  \nonumber
  \end{align}
} 
\begin{align}
  n(A,B) 
  &=
  \frac{1}{V_{d+1}}\,
  \sum_x\,\bigl[ (D_a^{-1})_{xx}\,(h_a)_x 
    - (D_b^{-1})_{xx}\,(h_b)_x \bigr]
  +1,
\label{number_density2}
\\
  e(A,B) 
  &=\frac{1}{V_{d+1}}\,\sum_x\,\biggl[
    \Bigl( \frac{U}{2} - \frac{i}{2}\,\sqrt{\frac{\alpha U}{\epsilon}}\,A_x \Bigr)\,
      \bigl[ (D_a^{-1})_{xx}\,(h_a)_x - (D_b^{-1})_{xx}\,(h_b)_x \bigr]
\nonumber
\\
  &~~~~
    + \Bigl(
        (1-\alpha)\,U - \frac{1}{2}\,\sqrt{\frac{(1-\alpha) U}{\epsilon}}\,B_x
      \Bigr)\,
      \bigl[ (D_a^{-1})_{xx}\,(h_a)_x + (D_b^{-1})_{xx}\,(h_b)_x \bigr]
\nonumber
\\
  &~~~~
    - \bigl[(t\,D^{-1}_a)_{x x}\,(h_a)_x + (t\,D^{-1}_b)_{x x}\,(h_b)_x \bigr]
  \biggr].
\label{energy_density2}
\end{align}
Note that for the hopping matrix \eqref{hopping}, 
the last line of Eq.~\eqref{energy_density2} is written as 
\begin{align}
    - t\,\sum_{i=1}^d\,\Bigl\{
      \bigl[ (D_a^{-1})_{x+i,x} + (D_a^{-1})_{x-i,x} \bigr]\,(h_a)_x
      + 
      \bigl[ (D_b^{-1})_{x+i,x} + (D_b^{-1})_{x-i,x} \bigr]\,(h_b)_x
    \Bigr\} ,
\end{align}
where $x \pm i$ denotes a positive/negative shift of the coordinate $x$ 
in the $i$-th direction

\section{Results on the one-dimensional Hubbard model}
\label{sec:results1}

We first verify the algorithmic correctness of WV-HMC 
in a simple setup with a mild sign problem.  
We consider a one-dimensional spatial lattice of size $L_s = 4$ 
at inverse temperature $\beta = 0.2$, 
with Trotter number $N_t = 4$ 
(corresponding to Trotter step $\epsilon = 0.05$). 
The hopping amplitude and interaction strength 
are set to $t = 1.0$ and $U = 4.0$, respectively. 
The shifted chemical potential $\tilde\mu = \mu - U/2$ 
is varied over the range $[-6.0, 6.0]$. 
The flow time interval is set to $[T_0, T_1] = [0.02, 0.10]$, 
with weight function parameters $\gamma = 0$, $c_0 = c_1 = 1$, 
and $d_0 = d_1 = 0.02$ [see Eq.~\eqref{Wt}]. 

We measure the number density $n$ [Eq.~\eqref{number_density2}] 
and the energy density $e$ [Eq.~\eqref{energy_density2}] 
using WV-HMC, 
with the redundant parameter $\alpha$ [Eq.~\eqref{alpha}]
set to two different values: $\alpha = 0.1$ and $\alpha = 1.0$. 
The exact values used for comparison are obtained 
using the method of Ref.~\cite{Fukuma:2019wbv}, 
which is designed to yield exact results for a finite Trotter number $N_t$. 
We also compare the results 
with those obtained using the complex Langevin (CL) method  
\cite{Parisi:1983cs,Klauder:1983sp}.

\begin{figure}[t]
  \centering
  \includegraphics[width=80mm]{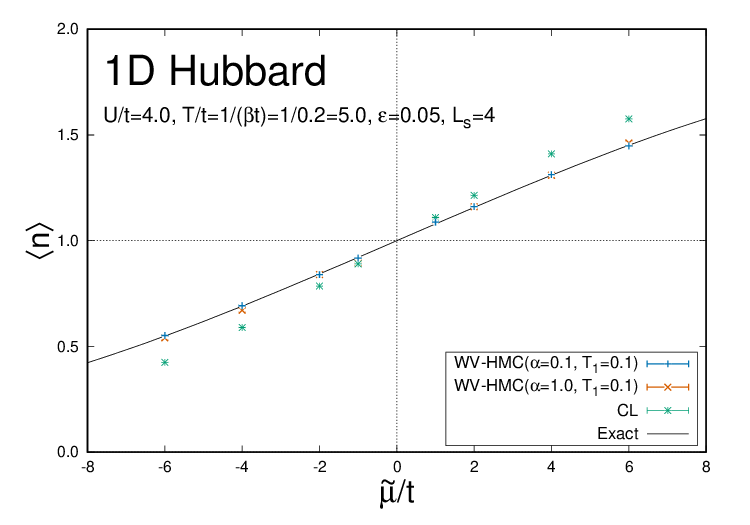}
  \includegraphics[width=80mm]{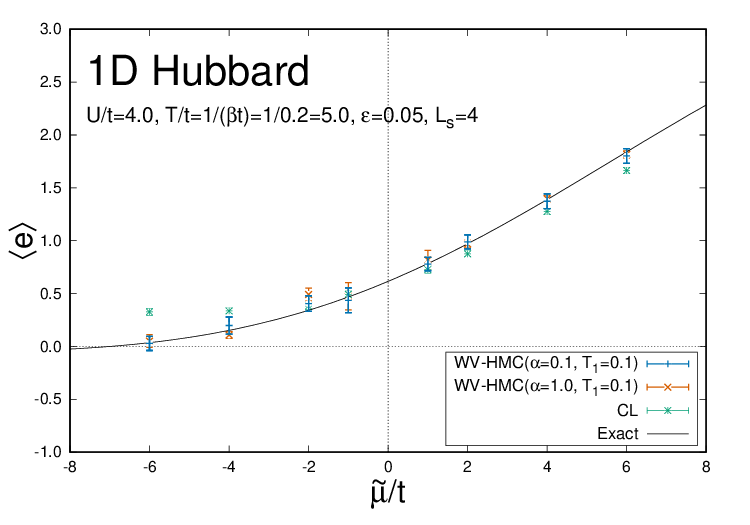}
  \caption{
    Number densities $\vev{n}$ (left) 
    and energy densities $\vev{e}$ (right) 
    obtained using WV-HMC 
    on a spacetime lattice of size $N_t \times L_s = 4 \times 4$ 
    (sample size: 200). 
    The parameters are set to $t=1.0$, $U=4.0$, and $\beta=0.2$, 
    with various values of $\tilde\mu$. 
    WV-HMC simulations are performed 
    with two different values of the redundant parameter $\alpha$: 
    $\alpha = 0.1$ and $\alpha = 1.0$.  
    Both values reproduce the exact results within statistical errors, 
    which are smaller for smaller $\alpha$ 
    (corresponding to a milder sign problem 
    on the original integration surface $\Sigma_0$). 
    For comparison, 
    we also show results obtained using the complex Langevin (CL) method 
    (sample size: 1,000), 
    which exhibit incorrect convergence. 
  }
\label{fig:number_energy_density_1D_hubbard}
\end{figure}

\begin{figure}[t]
  \centering
  \includegraphics[width=54mm]{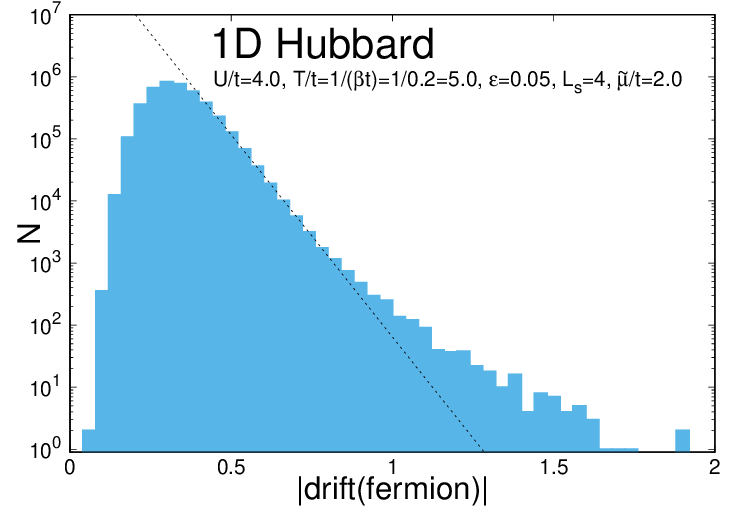}
  \includegraphics[width=54mm]{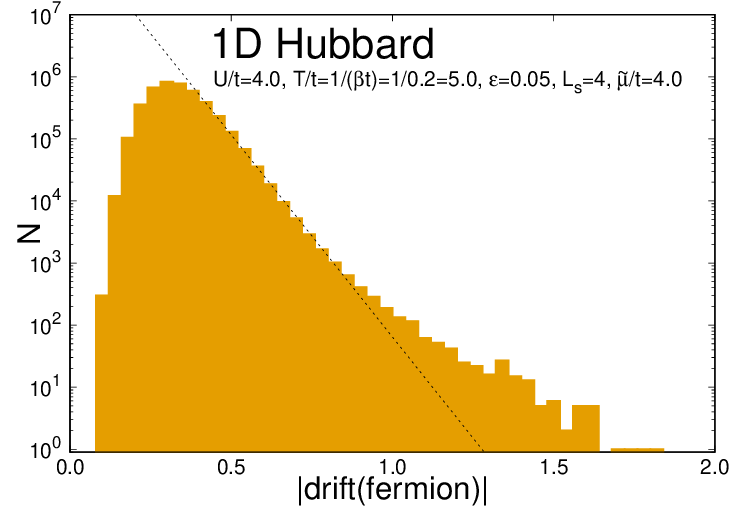}
  \includegraphics[width=54mm]{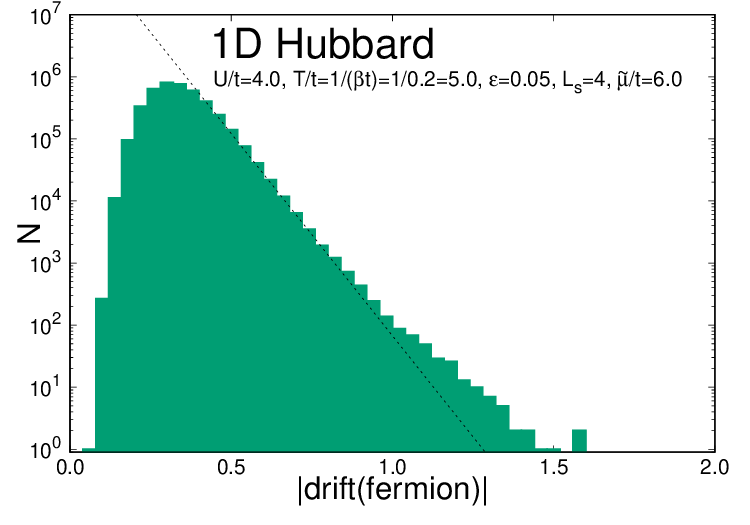}
  \caption{
    Histogram of the fermion drift norm in CL 
    on a spacetime lattice of size $N_t \times L_s = 4 \times 4$. 
    Parameters are set to $t=1.0$, $U=4.0$ and $\beta=0.2$, 
    with $\tilde\mu = 2.0$ (left), $\tilde\mu = 4.0$ (center), 
    and $\tilde\mu = 6.0$ (right).
  }
\label{fig:1D_hubbard_histogram_drift}
\end{figure}

Figure~\ref{fig:number_energy_density_1D_hubbard} 
shows the number and energy densities 
obtained by the above methods. 
We observe that the WV-HMC results reproduce the exact values 
and exhibit negligible dependence on the choice of $\alpha$. 
The figure also indicates that 
the CL method converges to incorrect limits, 
as pointed out in Ref.~\cite{Yamamoto:2015ura}. 
In fact, 
the histogram of the drift norm (Fig.~\ref{fig:1D_hubbard_histogram_drift})
shows a long tail at large drift values, 
which signals the breakdown of the CL method 
\cite{Nagata:2016vkn} 
(see also Refs.~\cite{Aarts:2009uq,Aarts:2011ax,
Nishimura:2015pba,Nagata:2015uga}).
%

\section{Results on the two-dimensional Hubbard model}
\label{sec:results2}

We perform simulations of the two-dimensional doped Hubbard model 
at low temperature, 
including regimes where the sign problem is severe.
Two spatial lattice volumes are considered: 
$L_s \times L_s = 6 \times 6$ and $8 \times 8$,
at inverse temperature $\beta = 6.4$ 
with Trotter number $N_t = 20$ 
(corresponding to Trotter step $\epsilon = 0.32$).
The hopping amplitude and interaction strength 
are set to $t = 1.0$ and $U = 8.0$, respectively, 
corresponding to $T/t = 1/6.4 \simeq 0.156$ 
and $U/t = 8.0$. 
The shifted chemical potential $\tilde\mu = \mu - U/2$ 
is varied over the range $[0.5, 9.0]$, 
which includes two plateau regions in $\vev{n}$ 
at small and large values of $\tilde\mu$. 
We compare our results 
with those obtained using ALF \cite{Bercx:2017pit,ALF:2020tyi}, 
a highly efficient non-thimble DQMC code, 
whose computational cost scales as $O(N_t \times V_d^3)$ 
when the sign problem is absent, 
but becomes exponential when the problem exists.

\subsection{Computational cost scaling}

\begin{figure}[t]
  \centering
  \includegraphics[width=80mm]
    {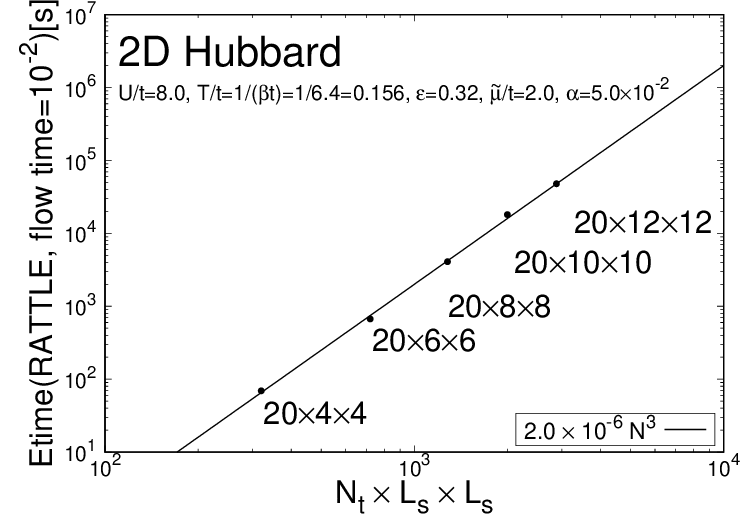}
  \caption{
    Elapsed time scaling of RATTLE update for various spacetime volumes.
  }
\label{fig:etime}
\end{figure}

Since we employ direct solvers for inverting the fermion matrices, 
the computational cost is expected to scale as $O(N^3)$, 
provided that the convergence rate of the Newton iteration 
in each RATTLE update 
(for solving the equation that determines the Lagrange multiplier $\lambda$) 
depends only weakly on the system size.  
Figure~\ref{fig:etime} shows the elapsed computational time per RATTLE update, 
clearly confirming this scaling behavior. 
The elapsed time is measured using GT-HMC rather than WV-HMC, 
because fixing the flow time yields more accurate timing measurements.
%

\subsection{Tuning of $\alpha$}

\begin{figure}[ht]
  \centering
  \includegraphics[width=70mm]{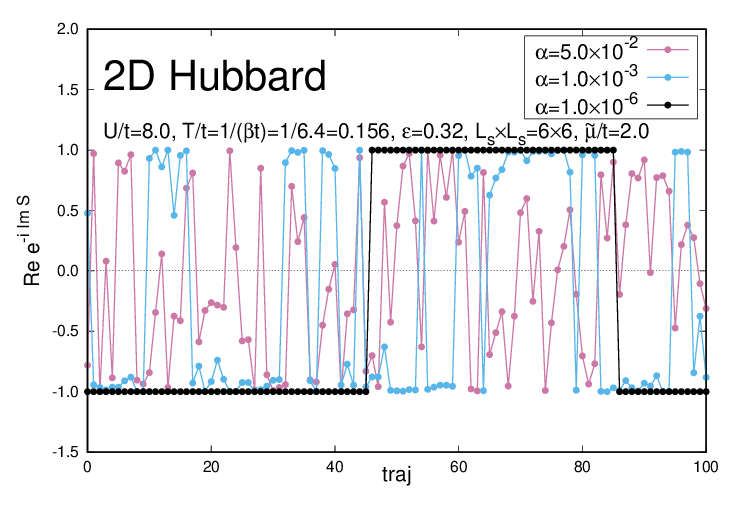}
  \includegraphics[width=70mm]{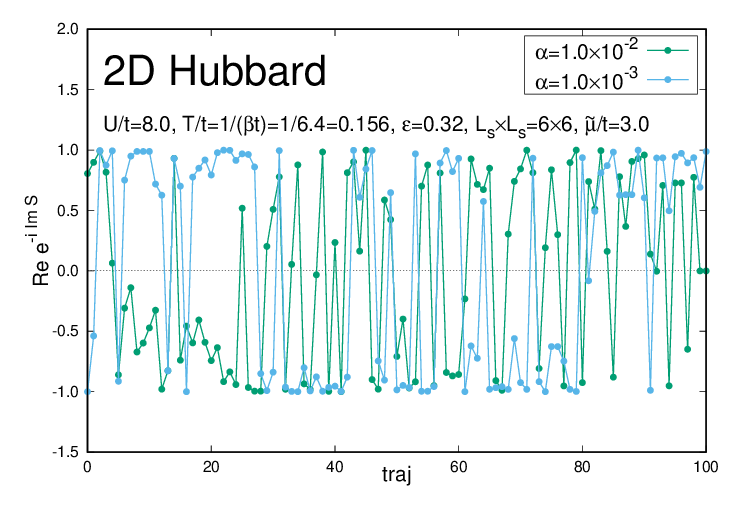}
  \includegraphics[width=70mm]{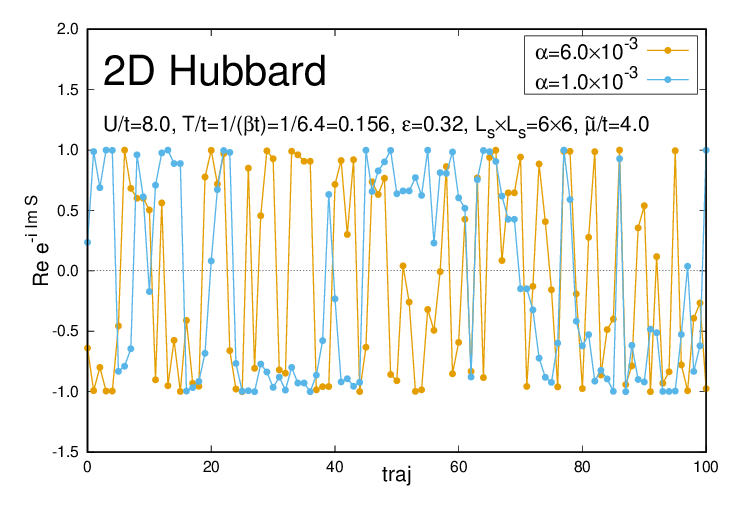}
  \includegraphics[width=70mm]{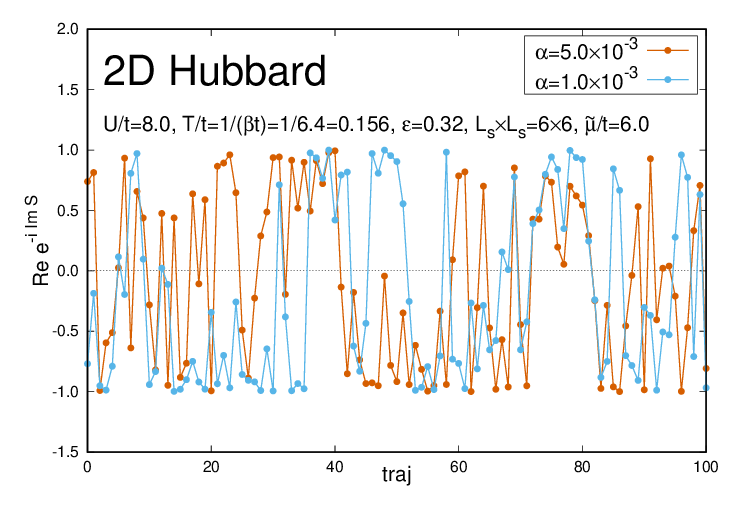}
  \caption{
    ($N_t \times L_s^2 = 20 \times 6 \times 6$) 
    History of the phase factor on $\Sigma_0$  
    for various values of $\alpha$ for $\tilde\mu = 2.0,\,3.0,\,4.0,\,6.0$ 
    (from top left to bottom right). 
  }
\label{fig:alpha_dependence}
\end{figure}

To reduce the computational cost, 
it is desirable to suppress the sign problem 
on the original integration surface $\Sigma_0$ 
as much as possible before starting simulations. 
In our algorithm, 
this can be achieved by choosing a small value of $\alpha$. 
However, care must be taken, 
because too small a value may lead to ergodicity issues
(see Sect.~\ref{sec:bosonization}) \cite{Beyl:2017kwp}. 
Figure~\ref{fig:alpha_dependence} shows 
the history of the phase factor $e^{-i\,\ImS}$ on $\Sigma_0$ 
for various values of $\alpha$ with $\tilde\mu = 2.0,\,3.0,\,4.0,\,6.0$ 
on a spacetime lattice of volume $N_t \times L_s^2 = 20 \times 6 \times 6$. 
We observe that 
the autocorrelation time (estimated from the average plateau length) 
increases as $\alpha$ decreases. 
This indicates that ergodicity issues become more severe, 
reflecting the presence of zeros of the fermion determinants on $\Sigma_0$. 
In our simulations, 
we adopt the following criterion for selecting $\alpha$: 
the length of any plateau must be shorter than 10 trajectories. 
The selected values of $\alpha$ are summarized in Table~\ref{table:alpha_tuned_WV}.
\begin{table}[ht]
  \centering
   \begin{tabular}{|c|c|c|c|c|c|c|c|c|} \hline
    $\tilde{\mu}$ 
    &  0.5   &  1.0 -- 1.8  
    &  2.0   &  2.2--3.0  &  3.5       
    &  4.0   &  4.5 -- 9.0
    \\ \hline
    $\alpha$  
    &  0.5   &  0.1  
    &  0.05  &  0.01      &  0.008
    &  0.006 &  0.005 
    \\ \hline
  \end{tabular}
  \caption{
    ($N_t \times L_s^2 = 20 \times 6 \times 6$) 
    Tuned values of $\alpha$ used in WV-HMC simulations.
  }
\label{table:alpha_tuned_WV}
\end{table}

\begin{figure}[ht]
  \centering
  \includegraphics[width=70mm]
    {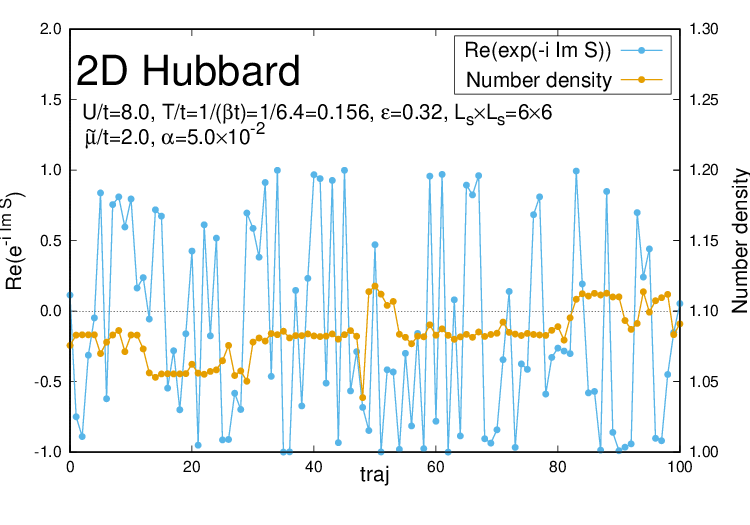}
  \hspace{5mm}
  \includegraphics[width=70mm]
    {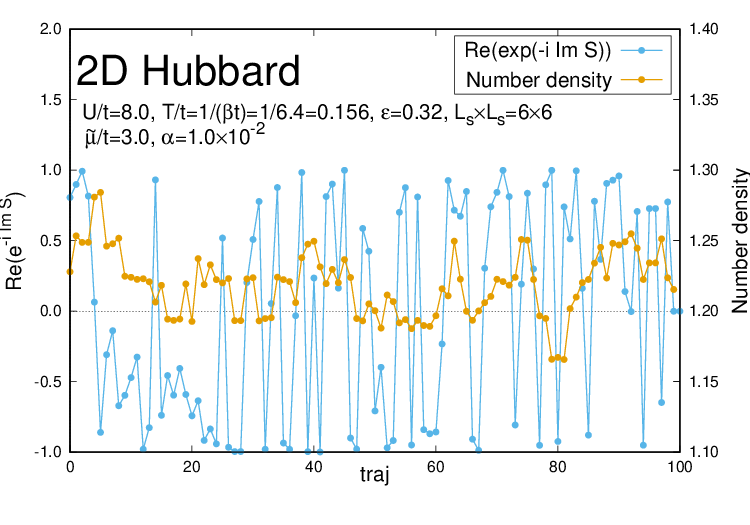}
  \\
  \includegraphics[width=70mm]
    {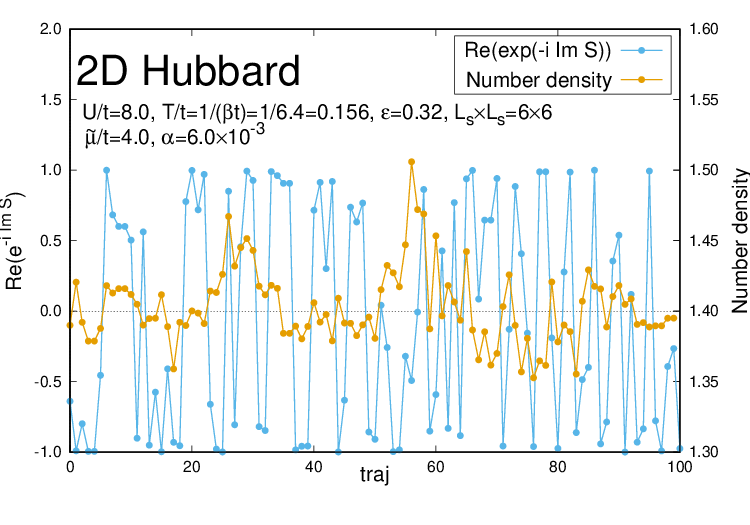}
  \hspace{5mm}
  \includegraphics[width=70mm]
    {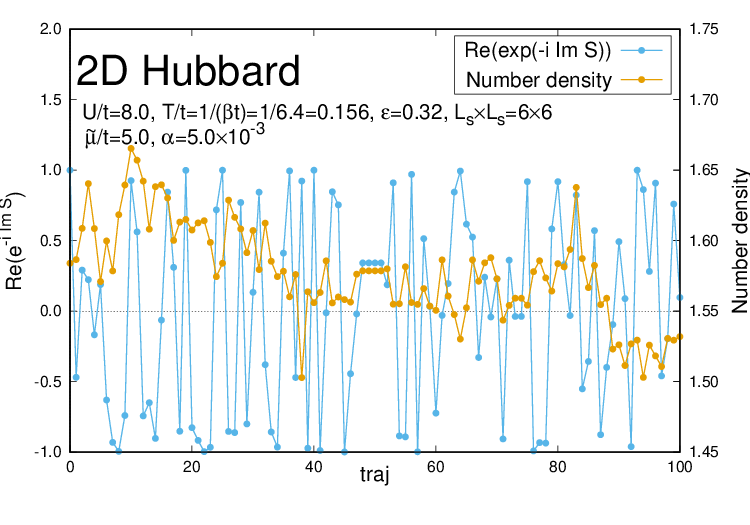}
  \\
  \includegraphics[width=70mm]
    {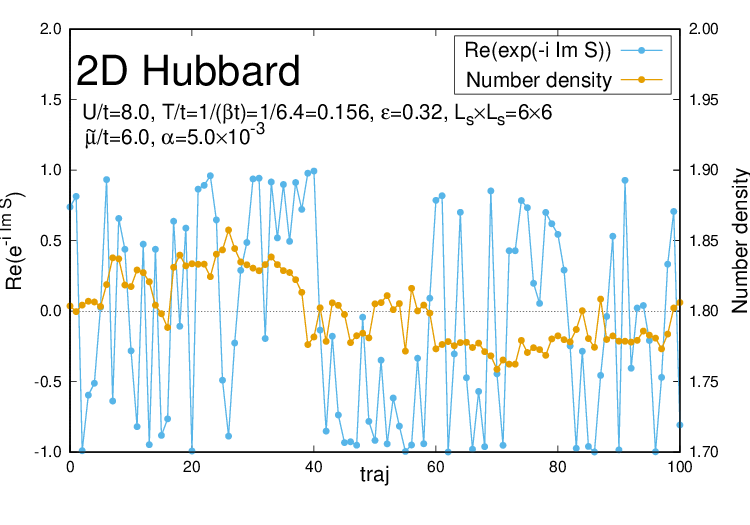}
  \hspace{5mm}
  \includegraphics[width=70mm]
    {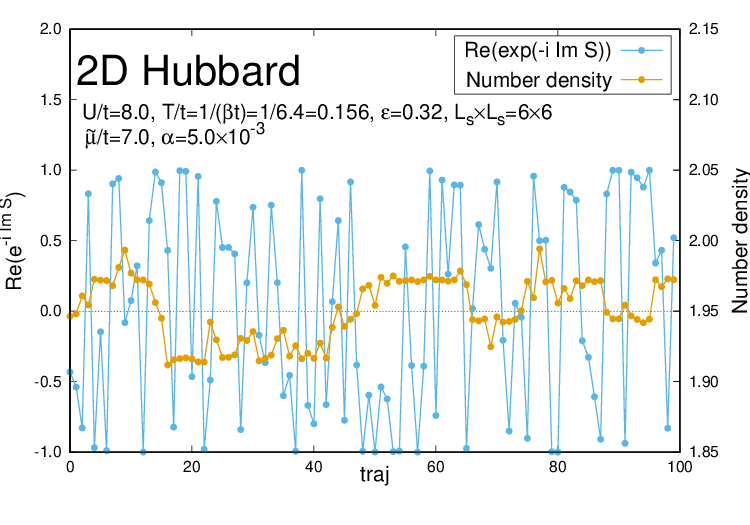}
  \\
  \includegraphics[width=70mm]
    {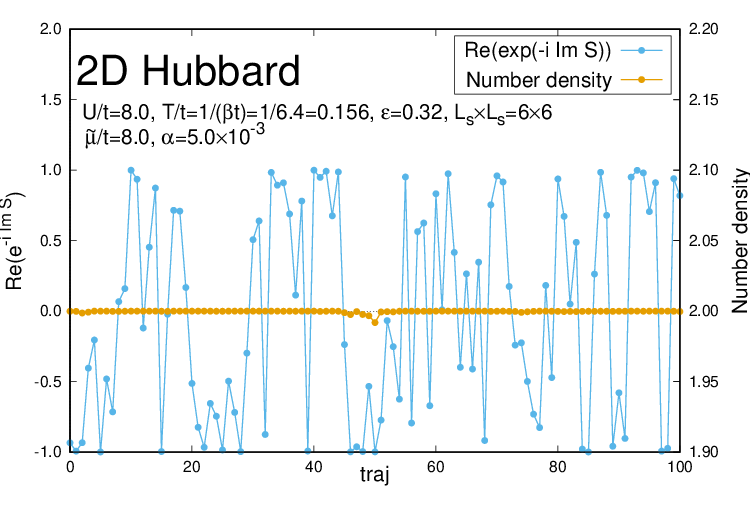}
  \hspace{5mm}
  \includegraphics[width=70mm]
    {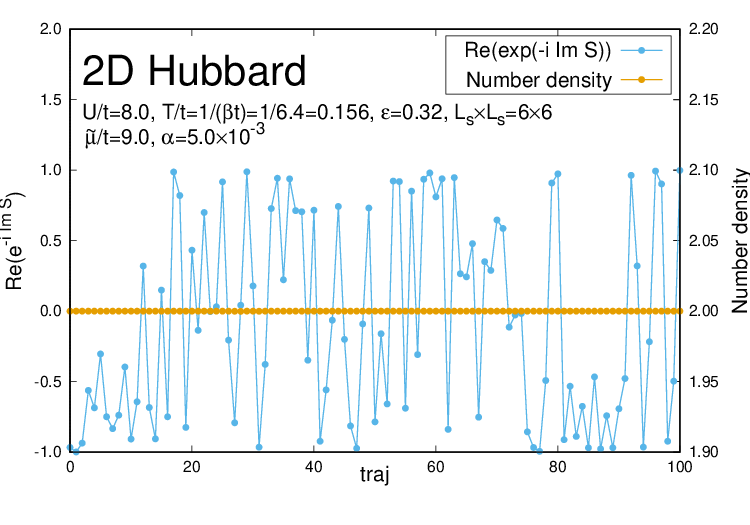}
  \caption{
    ($N_t \times L_s^2 = 20 \times 6 \times 6$) 
    Histories of the phase factor and the number density on $\Sigma_0$ 
    obtained using the tuned values of $\alpha$ in Table~\ref{table:alpha_tuned_WV}. 
    $\tilde\mu$ is varied from 2.0 to 9.0 (from top left to bottom right). 
  }
\label{fig:alpha_tuned}
\end{figure}

Figure~\ref{fig:alpha_tuned} shows 
the histories of both the phase factor and the number density on $\Sigma_0$
obtained using the tuned values of $\alpha$. 
The frequent fluctuations of the reweighting factor 
suggest that ergodicity issues are unlikely.
%

\subsection{Sign problem after $\alpha$ tuning}

\begin{figure}[ht]
  \centering
  \includegraphics[width=80mm]
    {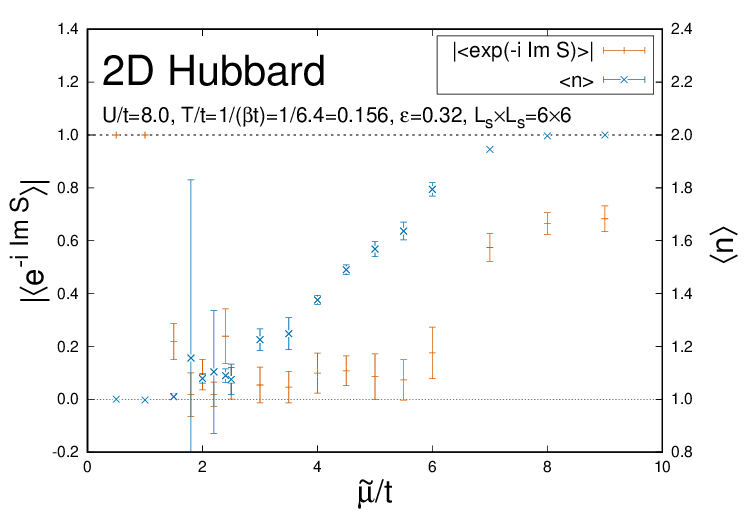}
  \caption{
    ($N_t \times L_s^2 = 20 \times 6 \times 6$) 
    Average phase factors and number densities on $\Sigma_0$ 
    at various values of $\tilde\mu$. 
  }
\label{fig:average_phase_factor_naiveRW}
\end{figure}

Figure~\ref{fig:average_phase_factor_naiveRW} shows 
the average phase factors on $\Sigma_0$ 
for a $20 \times 6 \times 6$ spacetime lattice, 
computed with the tuned values of $\alpha$ 
listed in Table~\ref{table:alpha_tuned_WV}. 
The figure demonstrates that 
while tuning $\alpha$ successfully resolves ergodicity issues on $\Sigma_0$, 
it does not fully remove the sign problem. 
Indeed, 
the average phase factors are statistically consistent with zero 
within one standard deviation 
in the range $1.8 \leq \tilde{\mu} \leq 5.5$.
The number densities are also shown in the same figure, 
exhibiting large statistical uncertainties 
that reflect the severity of the sign problem.

\begin{figure}[ht]
  \centering
  \includegraphics[width=70mm]{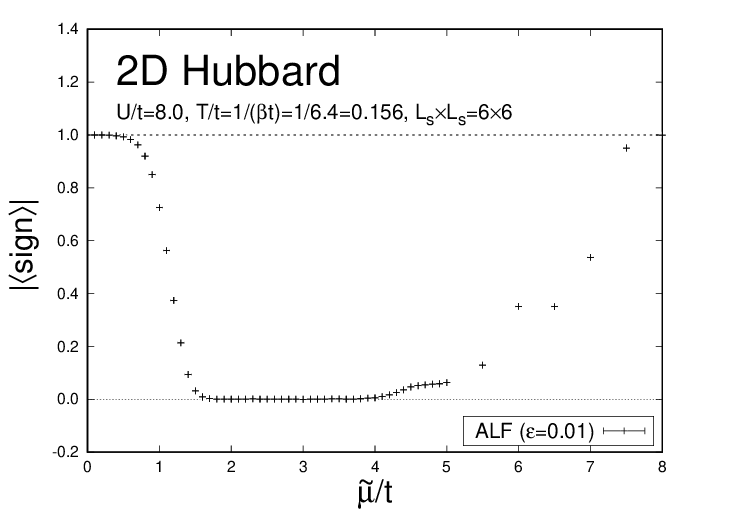}
  \includegraphics[width=70mm]{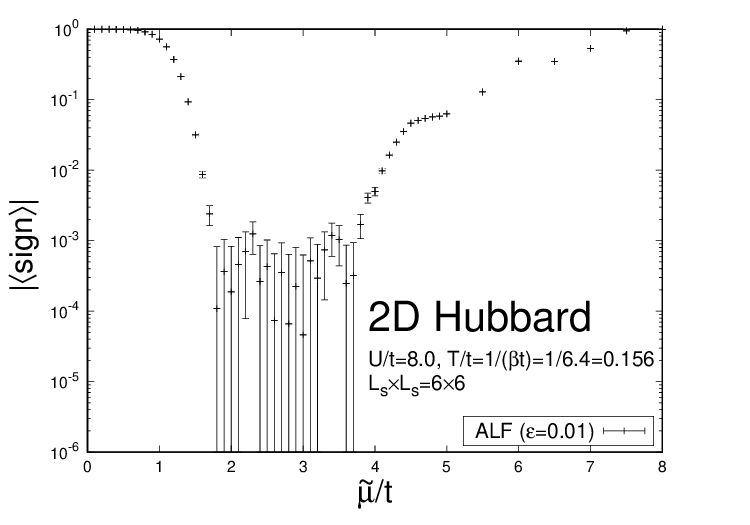}
  \caption{
    ($N_t \times L_s^2 = 20 \times 6 \times 6$) 
    Average signs at various values of $\tilde\mu$ obtained using ALF. 
  }
\label{fig:average_sign_factor_ALF_6x6}
\end{figure}

\begin{figure}[ht]
  \centering
  \includegraphics[width=70mm]{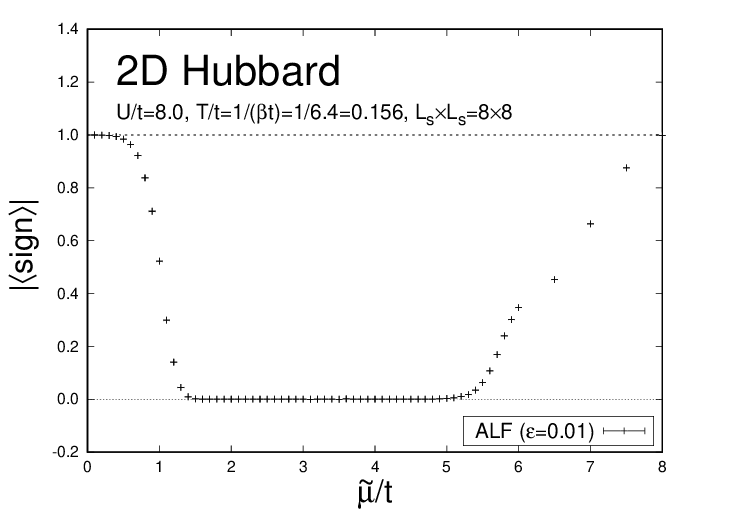}
  \includegraphics[width=70mm]{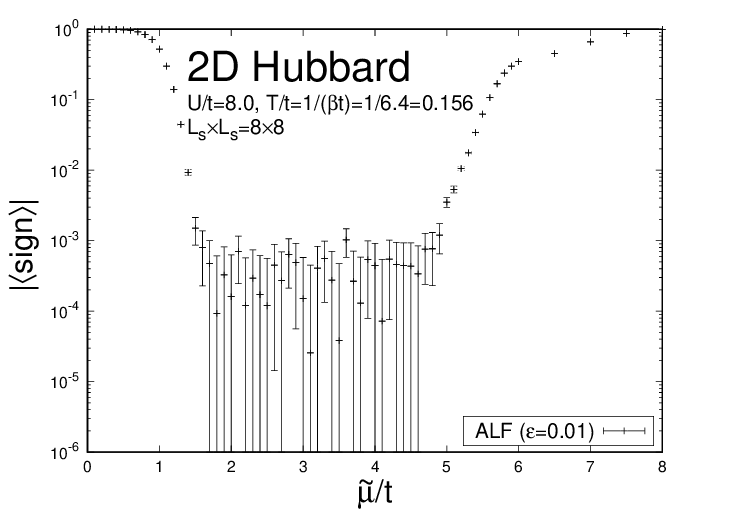}
  \caption{
    ($N_t \times L_s^2 = 20 \times 8 \times 8$) 
    Average signs at various values of $\tilde\mu$ obtained using ALF. 
  }
\label{fig:average_sign_factor_ALF_8x8}
\end{figure}

For comparison, 
Fig.~\ref{fig:average_sign_factor_ALF_6x6} presents 
results obtained using ALF 
on a $6 \times 6$ spatial lattice 
($\epsilon=0.01$; sample size: 50,000).
These results indicate that 
ALF also suffers from a severe sign problem 
in nearly the same parameter region.%
\footnote{ 
  Note that the exact coincidence of the two regions is not expected, 
  because different Hubbard-Stratonovich variables are used in ALF 
  ($Z_2$ variables in ALF, 
  while continuous Gaussian variables are used in our case).  
} 
Figure~\ref{fig:average_sign_factor_ALF_8x8} 
presents results obtained using ALF 
on an $8 \times 8$ spatial lattice 
($\epsilon=0.01$; sample size: 50,000).
We observe that the sign problem becomes more severe 
as the lattice volume increases. 
%

\subsection{Results on the $6 \times 6$ spatial lattice}
\label{sec:6x6}

In this subsection, 
we study a $6 \times 6$ spatial lattice 
at inverse temperature $\beta = 6.4$ 
with Trotter number $N_t = 20$. 
We also set $t = 1.0$ and $U = 8.0$. 
This parameter set corresponds to 
$T/t = 1/(t \beta) = 1/6.4 \simeq 0.156$ 
and $U/t = 8.0$. 

\begin{figure}[ht]
  \centering
  \includegraphics[width=70mm]{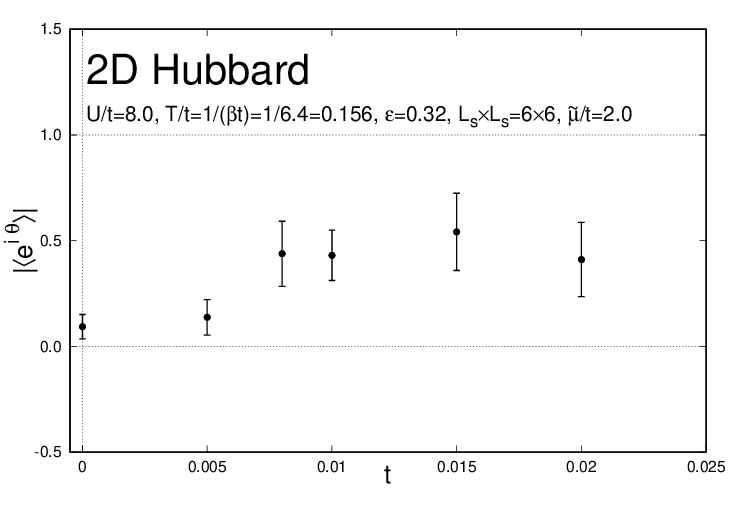}
  \includegraphics[width=70mm]{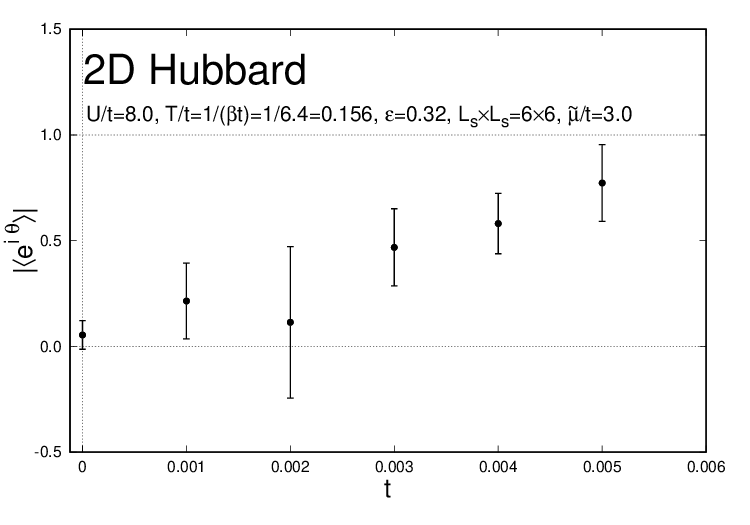}
  \includegraphics[width=70mm]{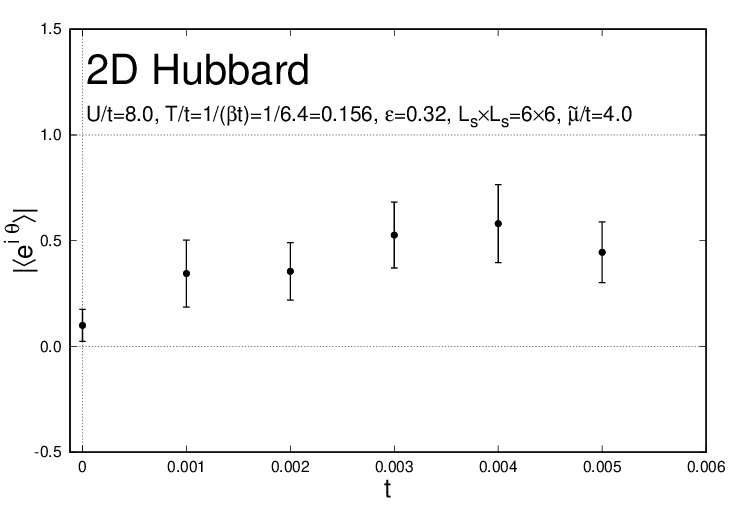}
  \includegraphics[width=70mm]{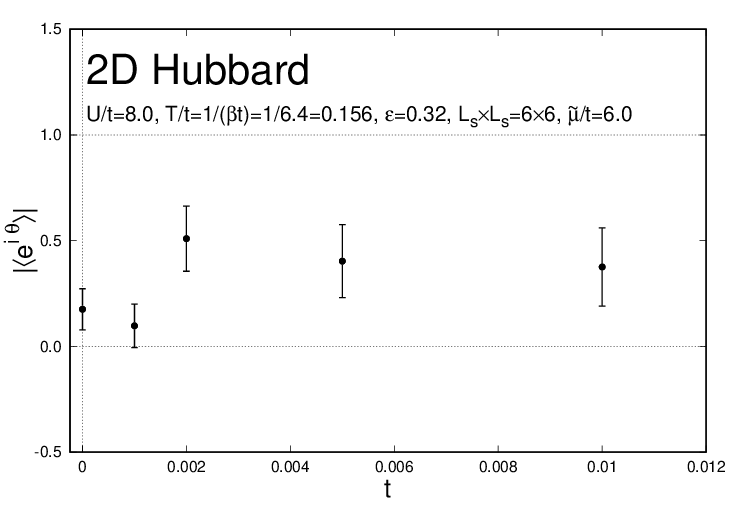}
  \caption{
    ($N_t \times L_s^2 = 20 \times 6 \times 6$) 
    Flow-time dependence of the average reweighting factor 
    $\vev{ e^{i \theta} }$ 
    for various values of $\tilde\mu$, 
    obtained using GT-HMC 
    [$e^{i\theta} \equiv (\det E / |\det E|) \, e^{-i\,\ImS}$].
  }
\label{fig:hubbard_T1}
\end{figure}

\begin{figure}[ht]
  \centering
  \includegraphics[width=70mm]{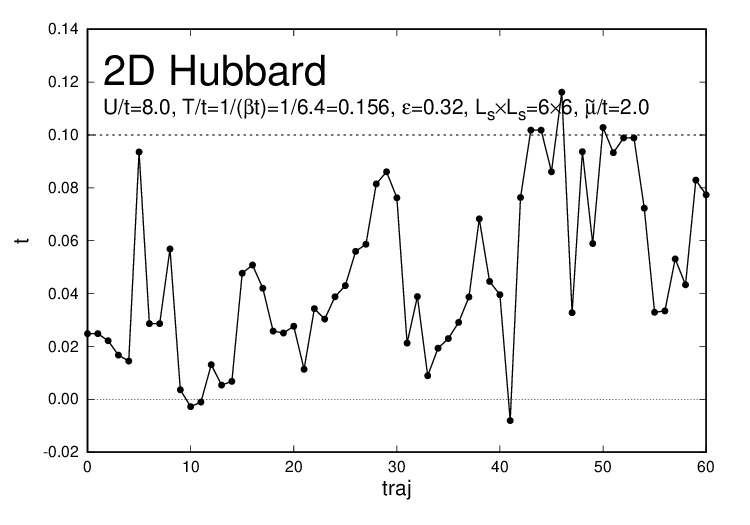}
  \includegraphics[width=70mm]{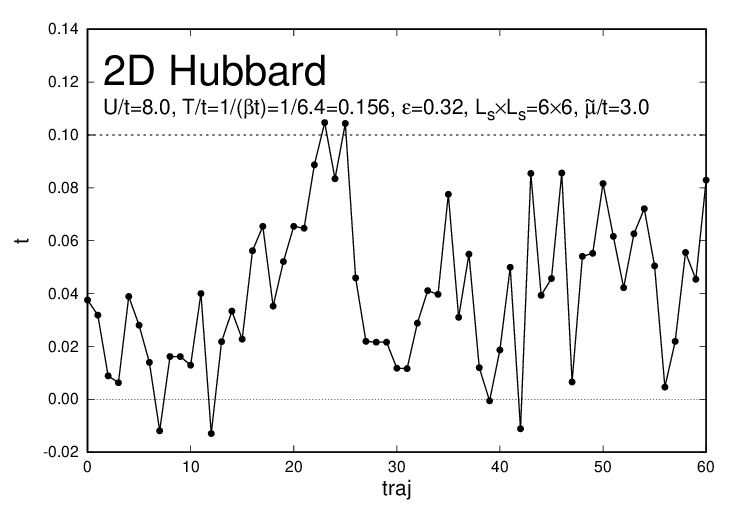}
  \includegraphics[width=70mm]{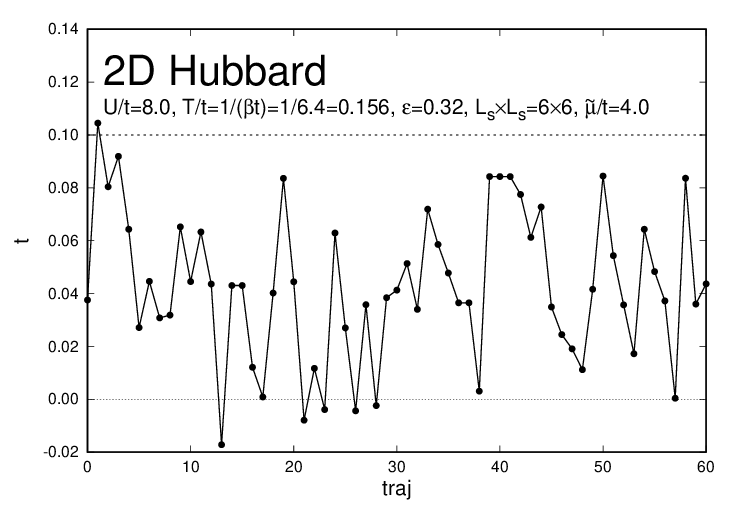}
  \includegraphics[width=70mm]{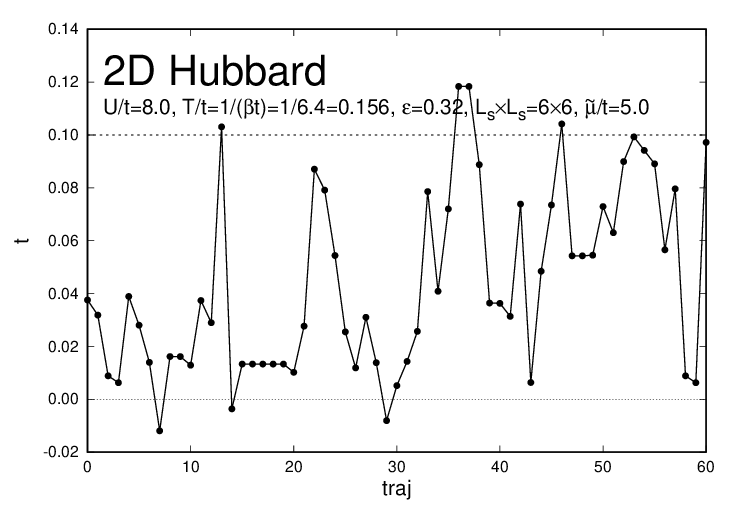}
  \includegraphics[width=70mm]{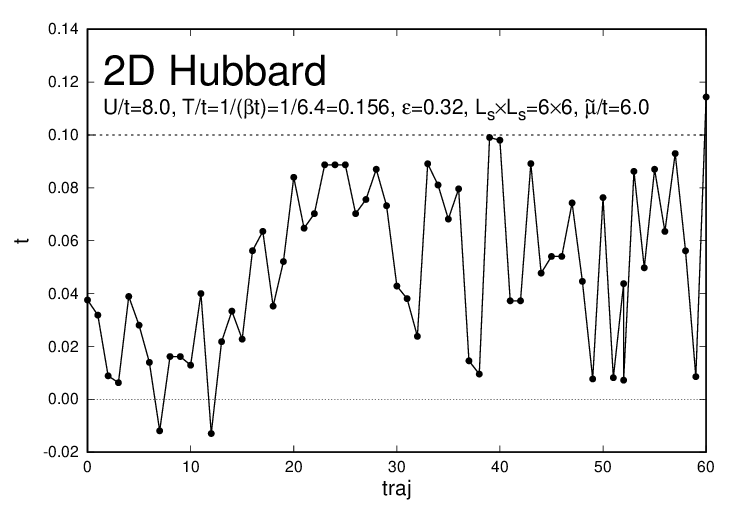}
  \includegraphics[width=70mm]{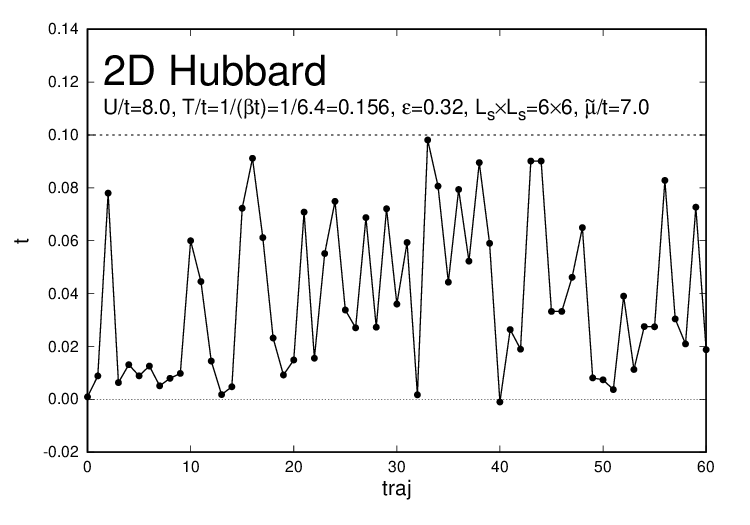}
  \includegraphics[width=70mm]{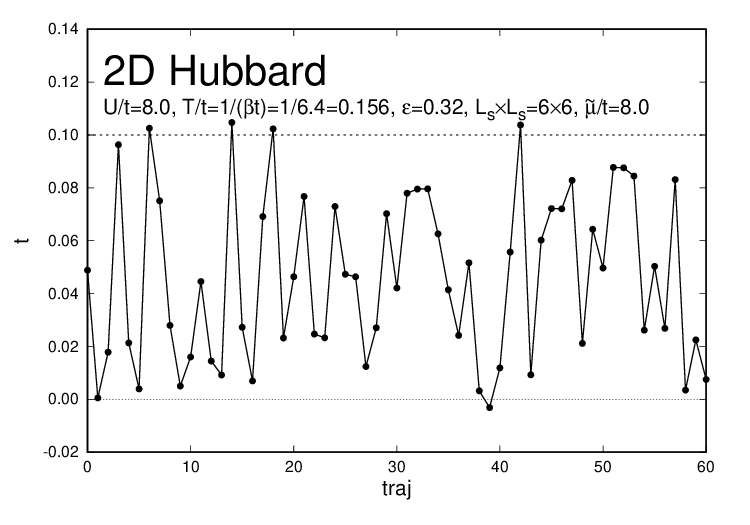}
  \includegraphics[width=70mm]{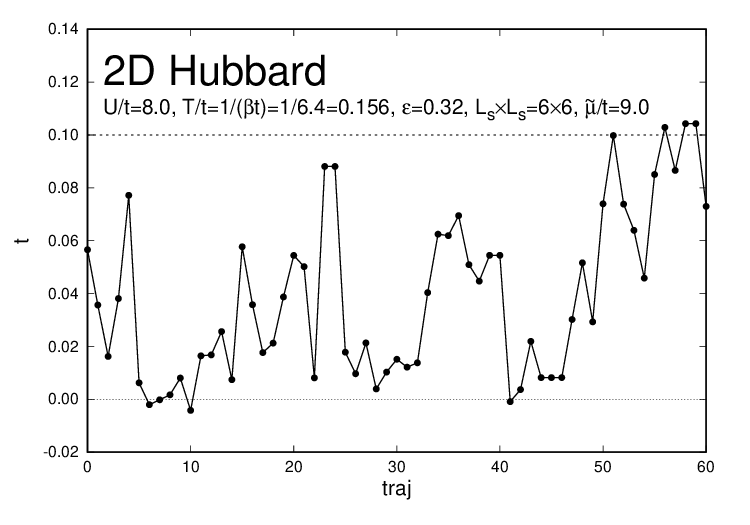}
  \caption{
    ($N_t \times L_s^2 = 20 \times 6 \times 6$) 
    History of the flow time in WV-HMC.
  }
\label{fig:tflow_6x6}
\end{figure}

Figure~\ref{fig:hubbard_T1} shows 
the average reweighting factor on the deformed surface $\Sigma_t$ 
at various flow times $t$, computed using GT-HMC, 
for $\tilde\mu = 2.0,\,3.0,\,4.0,\,6.0$. 
We observe that the average becomes 
statistically distinguishable from zero at the two-sigma level 
for $t > 10^{-2}$.
Based on these observations, 
we set the upper cutoff to $T_1 = 10^{-1}$ 
for the entire range of $\tilde\mu$ 
in the WV-HMC simulations.%
\footnote{ 
  The value $T_1 = 10^{-1}$ is in fact too large 
  for significantly reducing the sign problem. 
  However, choosing a smaller value 
  restricts the worldvolume to a thin layer, 
  for which configurations cannot be explored efficiently, 
  reintroducing ergodicity issues into the WV-HMC simulations. 
  A prescription for enhancing ergodicity in such a thin worldvolume 
  is to embed GT-HMC in WV-HMC, 
  as discussed in detail in Ref.~\cite{Fukuma:2025cxg}.
} 
The weight function parameters are chosen as 
$\gamma = 0$, 
$c_0 = c_1 = 0.01$, 
$d_0 = d_1 = 0.02$, 
with cutoffs $T_0 = 0.02$ and $T_1 = 0.10$.
Figure~\ref{fig:tflow_6x6} shows that 
the configurations efficiently explore the time interval $[T_0, T_1]$. 

\begin{figure}[ht]
  \centering
  \includegraphics[width=80mm]{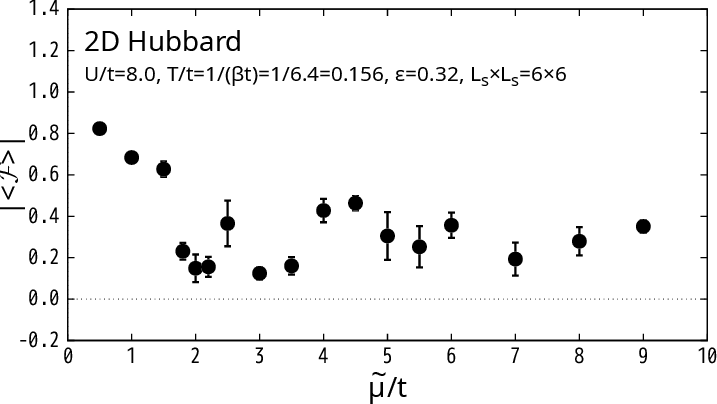}
  \caption{
    ($N_t \times L_s^2 = 20 \times 6 \times 6$) 
    Average reweighting factors obtained using WV-HMC.
  }
\label{fig:average_reweighting_factor_6x6}
\end{figure}

Figure~\ref{fig:average_reweighting_factor_6x6} 
shows the average reweighting factors $\vev{\calF}$ 
for various values of $\tilde\mu$.
Compared with the values on $\Sigma_0$  
in Fig.~\ref{fig:average_phase_factor_naiveRW}, 
these averages are significantly enhanced 
by incorporating configurations from $t > 0$ data. 

\begin{figure}[ht]
  \centering
  \includegraphics[width=125mm]
    {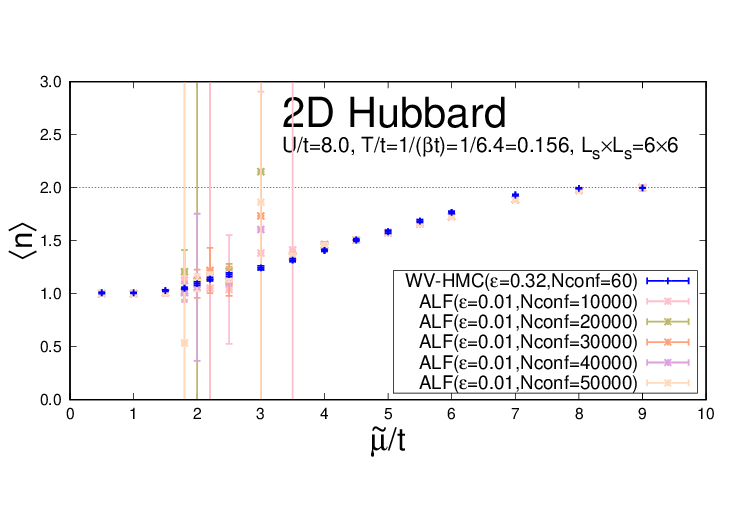}\\
  \vspace{-15mm}
  \includegraphics[width=125mm]
    {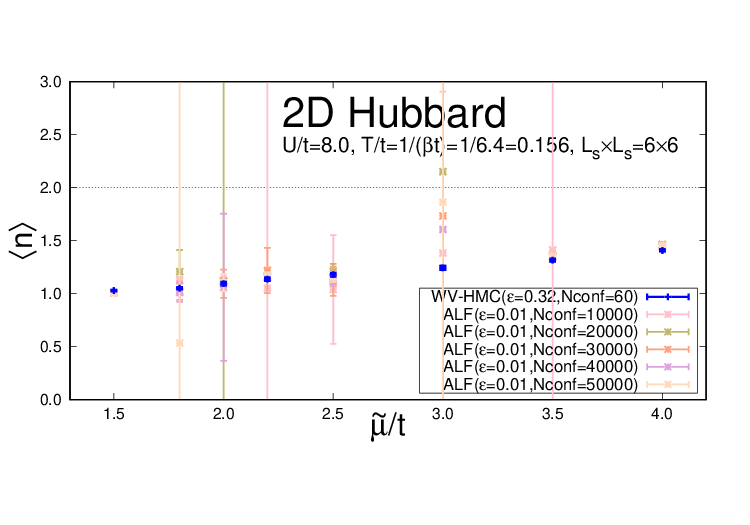}
  \caption{
    ($N_t \times L_s^2 = 20 \times 6 \times 6$) 
    Number densities obtained using WV-HMC. 
    ALF results are shown for comparison.
    Top: Full range of $\tilde\mu$. 
    Bottom: Enlarged view of $1.5 \leq \tilde\mu \leq 4.0$, 
    including ALF results with five different sample sizes.  
  }
\label{fig:number_density_6x6}
\end{figure}

\begin{figure}[ht]
  \centering
  \includegraphics[width=125mm]
    {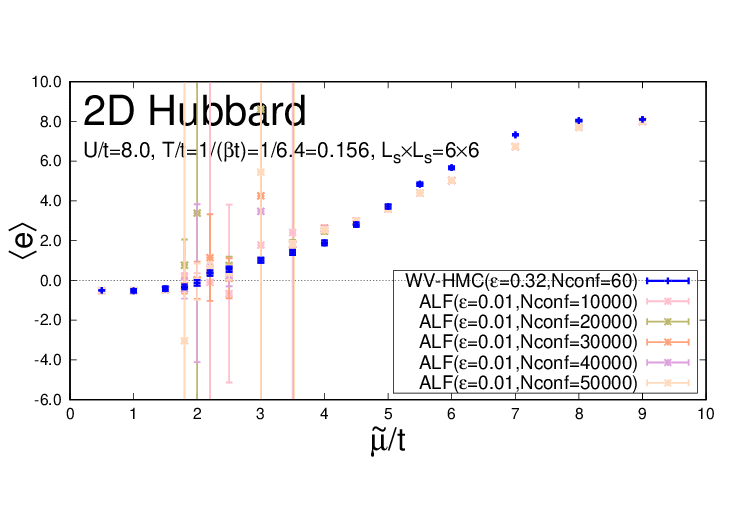}\\
  \vspace{-15mm}
  \includegraphics[width=125mm]
    {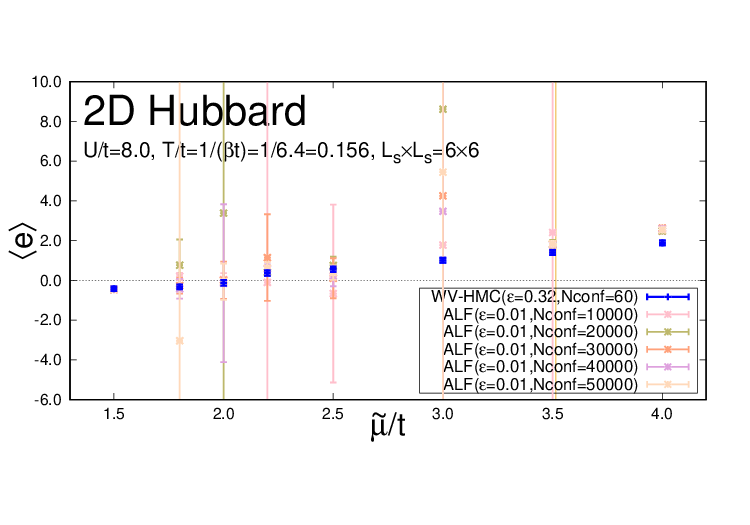}
  \caption{
    ($N_t \times L_s^2 = 20 \times 6 \times 6$) 
    Energy densities obtained using WV-HMC. 
    ALF results are shown for comparison.
    Top: Full range of $\tilde\mu$. 
    Bottom: Enlarged view of $1.5 \leq \tilde\mu \leq 4.0$, 
    including ALF results with five different sample sizes.  
  }
\label{fig:energy_density_6x6}
\end{figure}

Figures~\ref{fig:number_density_6x6} and \ref{fig:energy_density_6x6} 
show the number and energy densities obtained using WV-HMC. 
For comparison, 
the results obtained using ALF 
($\epsilon = 0.01$; sample size: 10,000--50,000)
are also included. 
We observe that 
WV-HMC yields results with small statistical uncertainties 
even in regions where the sign problem is severe, 
in sharp contrast to the results obtained using ALF.%
\footnote{ 
  A discrepancy in $\vev{e}$ between WV-HMC and ALF is observed 
  in the parameter region free from the sign problem. 
  We confirm that 
  this discrepancy originates from the finite Trotter step 
  and vanishes in the continuum limit $\epsilon \to 0$ \cite{Fukuma:2025cxg}.
} 
%

\subsection{Results on the $8 \times 8$ spatial lattice}
\label{sec:8x8}

\begin{figure}[ht]
  \centering
  \includegraphics[width=70mm]{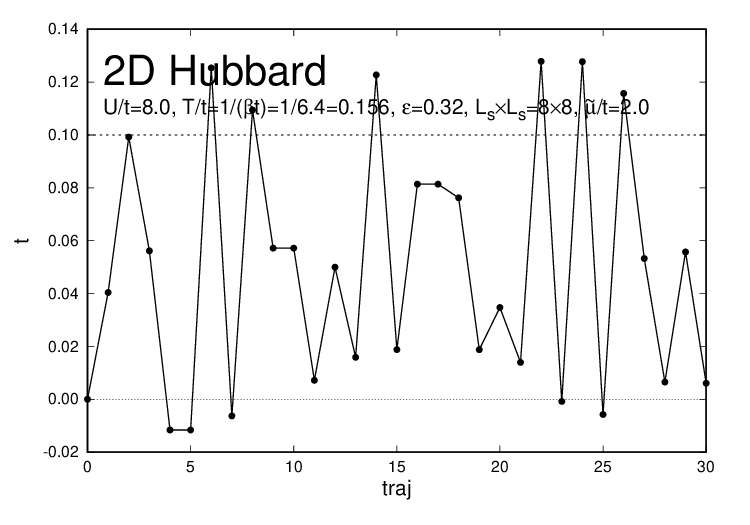}
  \includegraphics[width=70mm]{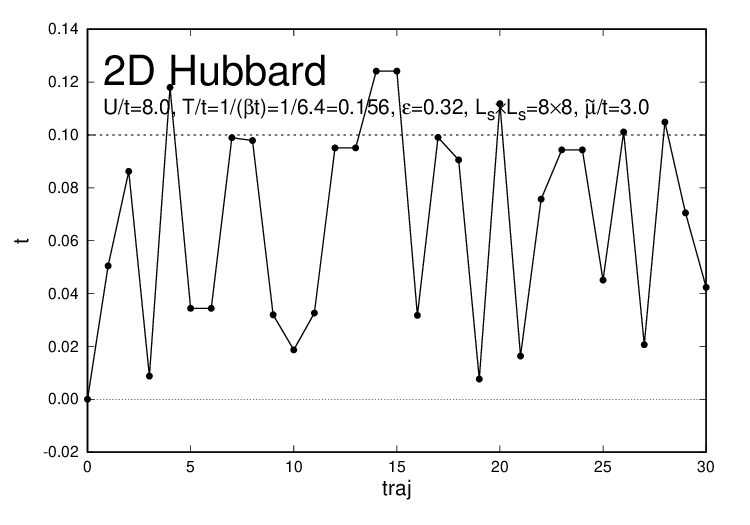}
  \includegraphics[width=70mm]{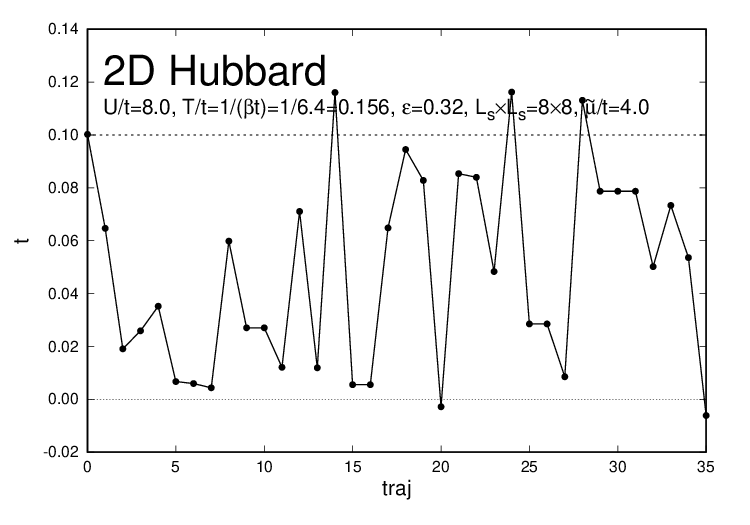}
  \includegraphics[width=70mm]{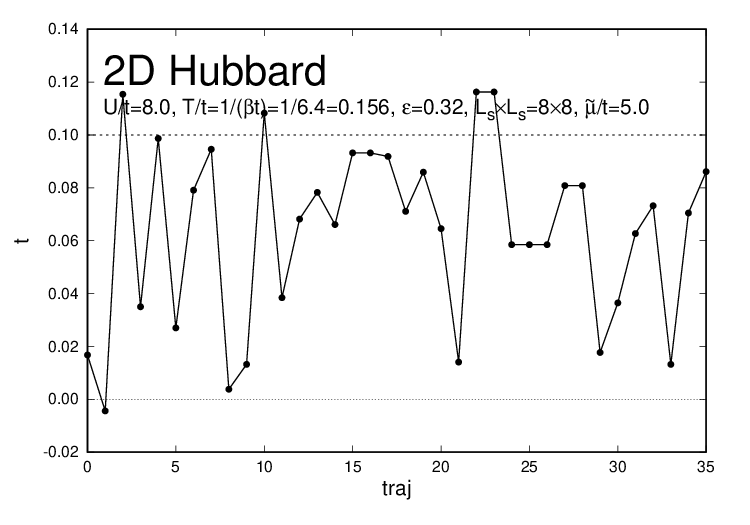}
  \includegraphics[width=70mm]{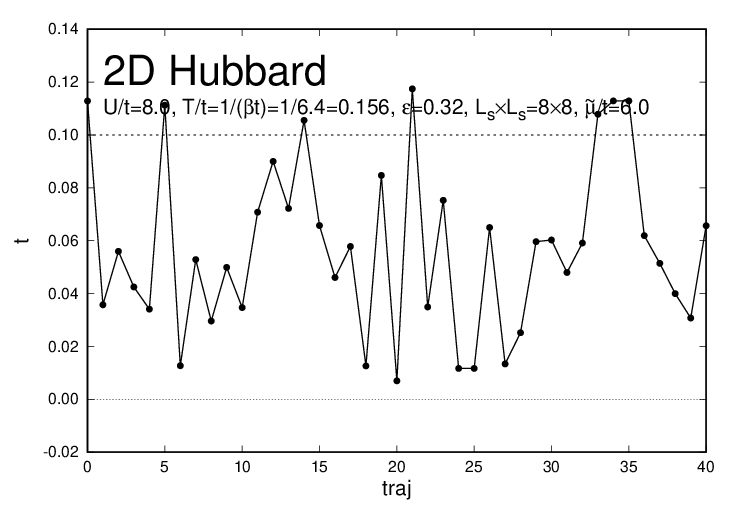}
  \includegraphics[width=70mm]{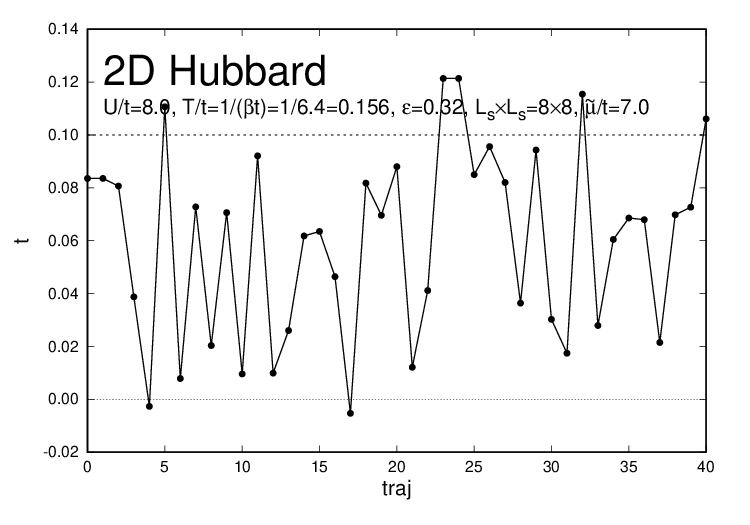}
  \includegraphics[width=70mm]{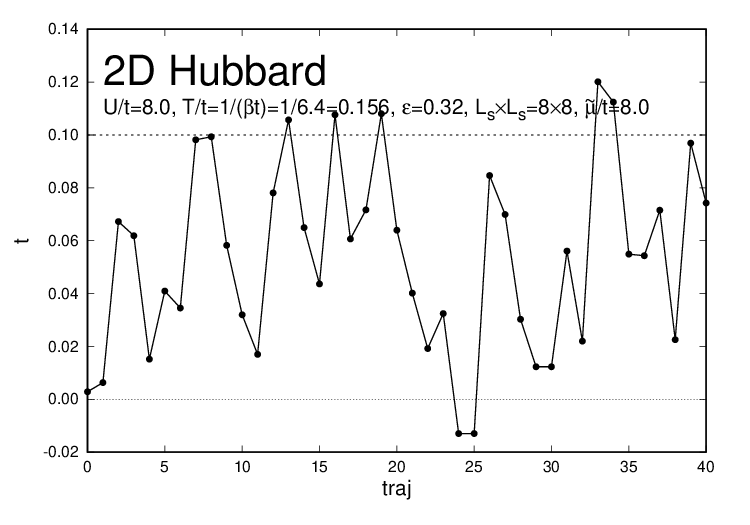}
  \includegraphics[width=70mm]{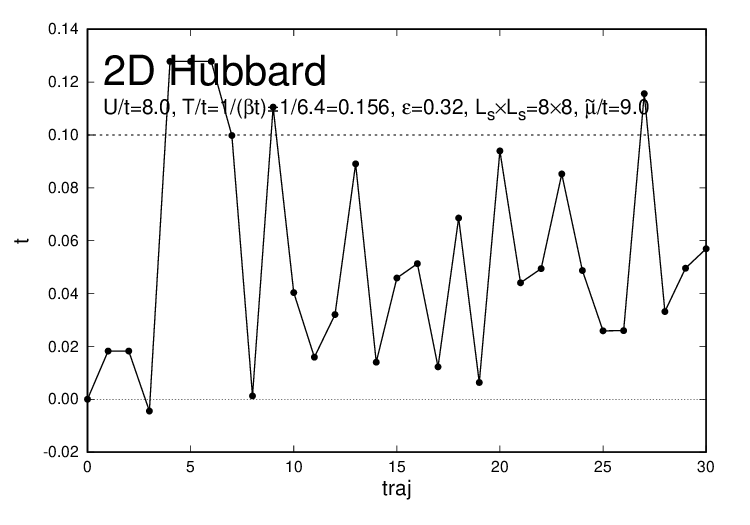}
  \caption{
    ($N_t \times L_s^2 = 20 \times 8 \times 8$) 
    History of the flow time in WV-HMC.
  }
  \label{fig:tflow_8x8}
\end{figure}

\begin{figure}[ht]
  \centering
  \includegraphics[width=80mm]{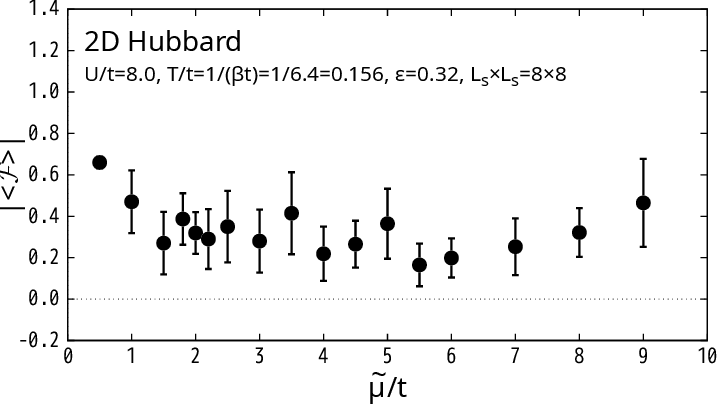}
  \caption{
    ($N_t \times L_s^2 = 20 \times 8 \times 8$) 
    Average reweighting factors obtained using WV-HMC.
  }
\label{fig:average_reweighting_factor_8x8}
\end{figure}

In this subsection, 
we consider an $8 \times 8$ spatial lattice 
at inverse temperature $\beta = 6.4$ 
with Trotter number $N_t = 20$. 
The simulation parameters are the same as those for the $6 \times 6$ case, 
except that we set $\gamma = 20$ 
to ensure nearly uniform sampling of the flow time 
within the interval $[T_0, T_1] = [0.02, 0.10]$.%
\footnote{ 
  A positive value of $\gamma$ is required for large lattices; 
  otherwise, the MD force 
  [which acts in the direction opposite to the flow; 
  see Eq.~\eqref{force}] 
  causes configurations to accumulate near the bottom of the worldvolume  
  (see Refs.~\cite{Fukuma:2020fez,Fukuma:2023eru} 
  for detailed explanations). 
  One can neglect this effect and set $\gamma = 0$ for small lattices 
  because the momentum refresh at the beginning of each MD trajectory 
  spreads configurations isotropically throughout the worldvolume.  
} 
Figure~\ref{fig:tflow_8x8} confirms that 
the entire region of flow time is sampled. 
Figure~\ref{fig:average_reweighting_factor_8x8} shows that 
the average reweighting factors are statistically nonvanishing 
at the one-sigma level. 

\begin{figure}[ht]
  \centering
  \includegraphics[width=125mm]
    {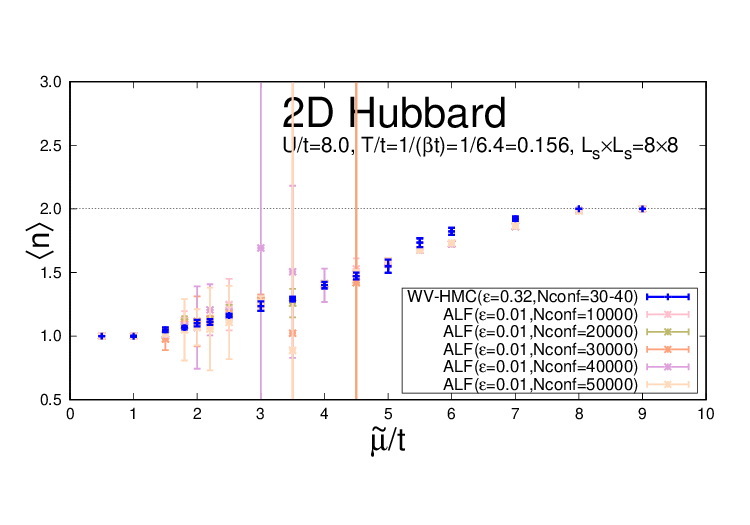}\\
  \vspace{-15mm}
  \includegraphics[width=125mm]
    {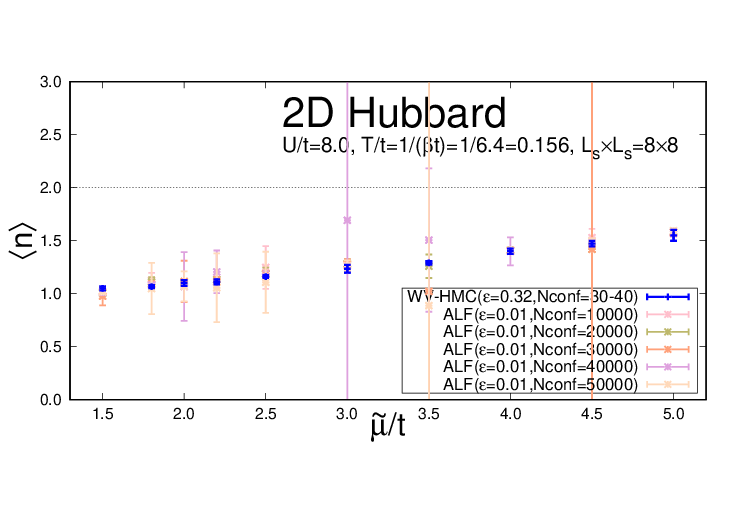}
  \caption{
    ($N_t \times L_s^2 = 20 \times 8 \times 8$) 
    Number densities obtained using WV-HMC. 
    ALF results are shown for comparison.
    Top: Full range of $\tilde\mu$. 
    Bottom: Enlarged view of $1.5 \leq \tilde\mu \leq 5.0$, 
    including ALF results with five different sample sizes.  
  }
\label{fig:number_density_8x8}
\end{figure}

\begin{figure}[ht]
  \centering
  \includegraphics[width=125mm]
    {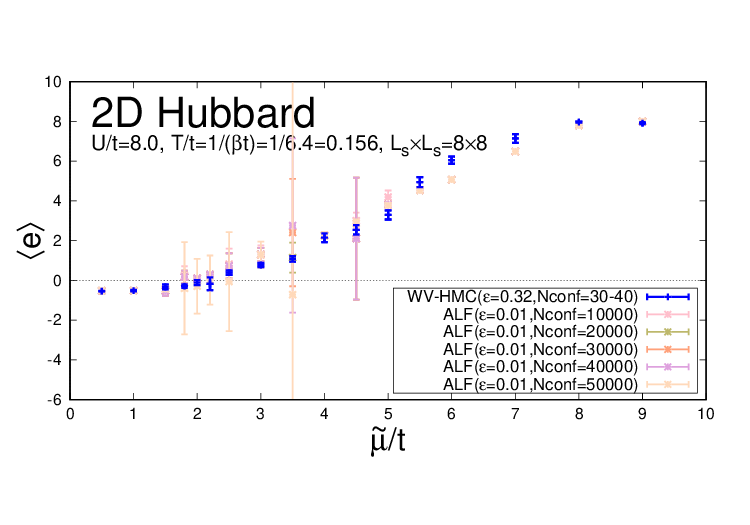}\\
  \vspace{-15mm}
  \includegraphics[width=125mm]
    {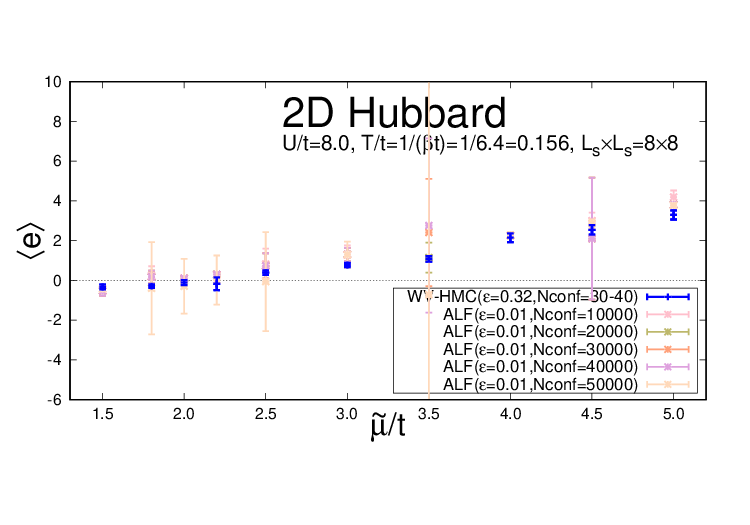}
  \caption{
    ($N_t \times L_s^2 = 20 \times 8 \times 8$) 
    Energy densities obtained using WV-HMC. 
    ALF results are shown for comparison.
    Top: Full range of $\tilde\mu$. 
    Bottom: Enlarged view of $1.5 \leq \tilde\mu \leq 5.0$, 
    including ALF results with five different sample sizes.  
  }
\label{fig:energy_density_8x8}
\end{figure}

Figures~\ref{fig:number_density_8x8} and \ref{fig:energy_density_8x8} 
present the number and energy densities 
obtained using WV-HMC. 
We also include the results obtained using ALF for comparison 
($\epsilon = 0.01$; sample size: 10,000--50,000).
Although the number of configurations is currently limited, 
the WV-HMC results remain statistically robust 
across the entire parameter range, 
as in the $6 \times 6$ case. 
%

\section{Conclusions and outlook}
\label{sec:conclusions}

In this paper, 
we applied the Worldvolume Hybrid Monte Carlo (WV-HMC) algorithm 
\cite{Fukuma:2020fez} 
to the two-dimensional doped Hubbard model. 
We used direct solvers for the inversion of fermion matrices 
and confirmed that the computational cost scales as $O(N^3)$, 
in agreement with theoretical expectations.
We evaluated the number density $\vev{n}$  
and the energy density $\vev{e}$ 
on $6 \times 6$ and $8 \times 8$ lattices 
at $T/t = 1/(t\beta) = 1/6.4 \simeq 0.156$ and $U/t = 8.0$ 
with Trotter number $N_t = 20$, 
and demonstrated that the WV-HMC method remains efficient,  
with well-controlled statistical errors  
even in parameter regions 
where the standard DQMC methods fail due to severe sign problems.

Although the computational cost of $O(N^3)$ 
remains high for practical calculations, 
especially when approaching the thermodynamic limit, 
our results suggest that 
the WV-HMC framework can serve as a powerful tool 
for investigating the doped Hubbard model. 
In fact, the cost can be reduced to $O(N^2)$ 
by introducing pseudofermions 
and employing iterative solvers \cite{Fukuma:2025esu}. 
However, this approach requires careful parameter tuning, 
a detailed study of which will be presented in a separate publication.

Even at finite spatial volumes, 
it is important to take the continuum limit in the temporal direction 
($\epsilon \to 0$), 
because the Trotter step $\epsilon$ used in the present paper 
is still relatively large ($\epsilon = 0.32$), 
although the code is designed 
such that finite-size corrections in the observables are $O(\epsilon^2)$. 
In the subsequent paper \cite{Fukuma:2025cxg}, 
we take the continuum limit 
after developing a method that enables us to reduce the maximum flow time 
while avoiding ergodicity issues in such a thin worldvolume. 

To investigate the model in the ground-state regime, 
one must further extrapolate to the $\beta \to \infty$ limit, 
using sufficiently large values of $\beta$ 
that already realize the $\epsilon\to 0$ limit. 
It would then be highly interesting 
to compare the zero-temperature Monte Carlo results 
(obtained using WV-HMC) 
with those from other approaches, 
such as variational Monte Carlo, 
constrained-path auxiliary-field quantum Monte Carlo, 
and density functional theory. 
One of the most informative indicators for such a comparison, 
which we plan to adopt, 
is the \emph{V-score} \cite{Wu:2023fgp}, 
a benchmark based on the variance of the ground-state energy.%
\footnote{ 
  We thank Masatoshi Imada for suggesting this comparison scheme. 
} 

In parallel with the present study, 
we are extending the WV-HMC method to other systems. 
An extension to group manifolds has already been completed 
\cite{Fukuma:2025gya}.
Current targets include finite-density QCD, 
other strongly correlated electron systems, 
and real-time dynamics of quantum many-body systems.
Results from these ongoing efforts will be reported in future publications.

\section*{Acknowledgments}

The authors thank 
Sinya Aoki, Fakher F.~Assaad, Masatoshi Imada, 
Ken-Ichi Ishikawa, Issaku Kanamori, Yoshio Kikukawa, 
Nobuyuki Matsumoto, Yusuke Nomura, Maksim Ulybyshev, 
Youhei Yamaji, and Shiwei Zhang 
for valuable discussions. 
M.F.\ acknowledges that parts of the basic lattice-field infrastructure of the code 
were developed using the FermiQCD/MDP library \cite{DiPierro:2001yu}. 
The WV-HMC algorithmic components, 
together with the configuration generation, random-number generation, 
fermion-matrix construction and solvers, 
and routines for observable measurements, were implemented independently by M.F.
This work was partially supported by JSPS KAKENHI
(Grant Numbers JP20H01900, JP21K03553, JP23H00112, JP23H04506, JP24K07052, JP25H01533);
by MEXT as 
``Program for Promoting Researches on the Supercomputer Fugaku'' 
(Simulation for basic science: approaching the new quantum era,
JPMXP1020230411);
and by SPIRIT2 2025 of Kyoto University.



\baselineskip=0.9\normalbaselineskip



\end{document}